\documentclass[sigconf]{acmart}

\renewcommand\footnotetextcopyrightpermission[1]{}
\renewcommand\acmConference[1]{}
\newcommand{\ignore}[1]{}

\usepackage{mathptmx} 

\usepackage{xspace}
\newcommand{\name}{\mbox{CoSpec}\xspace}

\usepackage{balance}
\usepackage{float}
\usepackage{xcolor}
\usepackage{pifont}
\usepackage{algorithm}
\usepackage{algorithmicx}
\usepackage{algpseudocode}
\usepackage{enumitem}
\usepackage{paralist}
\usepackage{multirow}
\usepackage{tabularx}
\usepackage{tablefootnote}
\usepackage{booktabs}
\usepackage{subfig}

\ignore{
\usepackage[normalem]{ulem}
\usepackage[sort,nocompress]{cite}

}

\usepackage{todonotes}
\ignore{

}

\ignore{

\AtBeginDocument{%
  \providecommand\BibTeX{{%
    \normalfont B\kern-0.5em{\scshape i\kern-0.25em b}\kern-0.8em\TeX}}}

	\copyrightyear{2019} 
	\acmYear{2019} 
	\acmConference[MICRO-52]{The 52nd Annual IEEE/ACM International Symposium on Microarchitecture}{October 12--16, 2019}{Columbus, OH, USA}
	\acmBooktitle{The 52nd Annual IEEE/ACM International Symposium on Microarchitecture (MICRO-52), October 12--16, 2019, Columbus, OH, USA}
	\acmPrice{15.00}
	\acmDOI{10.1145/3352460.3358279}
	\acmISBN{978-1-4503-6938-1/19/10}

\copyrightyear{2019}
\acmYear{2019}
\setcopyright{rightsretained}

\acmConference[MICRO'19]{The 52nd Annual IEEE/ACM International Symposium on Microarchitecture}{October 12-16, 2019}{Columbus, OH, USA}
\acmDOI{10.1145/XXXXX.XXXXXX}
\acmISBN{XXX}
}

\begin{document}

\title{Compiler Directed Speculative Intermittent Computation}

\author{Jongouk Choi}
\affiliation{%
  \institution{Purdue University}
}
\email{choi658@purdue.edu}

\author{Qingrui Liu}
\affiliation{%
  \institution{Xilinx}
}
\email{qingruil@xilinx.com}

\author{Changhee Jung}
\affiliation{%
  \institution{Purdue University}
}
\email{chjung@purdue.edu}

\begin{abstract}
\label{sec:abs} 
Energy harvesting systems have emerged as an alternative to
battery-operated embedded devices. 
Due to the intermittent nature of energy harvesting, researchers
equip the systems with nonvolatile memory (NVM) and crash consistency mechanisms.
\ignore{
However, prior works require non-trivial hardware modifications
, e.g., voltage monitoring system, extra energy buffer, nonvolatile flip-flops, dependence
tracking logic for identifying idempotence violation, and nonvolatile scratchpad.
Unfortunately, they all cause significant area/power/manufacturing
costs.
}
However, prior works require non-trivial hardware modifications, e.g., a voltage
monitor, nonvolatile flip-flops/scratchpad, dependence tracking modules, etc., thereby causing significant
area/power/manufacturing costs.

For low-cost yet performant intermittent computation, this paper presents
\name, a new architecture/compiler co-design scheme that works for commodity
in-order processors used in energy-harvesting systems.  To achieve crash
consistency without requiring unconventional architectural support, \name
leverages speculation assuming that power failure is not going to occur and
thus holds all committed stores in a store buffer (SB)---as if they were
speculative---in case of mispeculation.  \name compiler first partitions a
given program into a series of recoverable code regions with the SB size in
mind, so that no region overflows the SB.  When the program control reaches the
end of each region, the speculation turns out to be successful, thus releasing
all the buffered stores of the region to NVM.  If power failure occurs during the execution
of a region, all its speculative stores disappear in the volatile SB, i.e.,
they never affect program states in NVM.  Consequently, the interrupted region
can be restarted with consistent program states in the wake of power failure. 

To hide the latency of the SB release---i.e., essentially NVM writes---at
each region boundary, \name overlaps the NVM writes of the current region with the
speculative execution of the next region. Such instruction level parallelism
gives an illusion of out-of-order execution on top of the in-order processor,
achieving a speedup of more than 1.2X when there is no power outage.  Our
experiments on a set of real energy harvesting traces with frequent outages
demonstrate that \name outperforms the state-of-the-art scheme by 1.8$\sim$3X
on average.

\end{abstract}




\maketitle
\section{Introduction} 
\label{sec:intro} 
Energy harvesting systems continue to grow at a rapid pace. The range of their
applications is becoming wider including IoT devices, vehicle tire pressure
sensors, health monitoring applications, unmanned vehicles, and so on
~\cite{hglee14,wslee14,hester2016towards,yildirim2016synchronization,nwafor2017towards,nirjon2018lifelong,sisinni2018industrial}.
The biggest challenge of energy harvesting systems is the instability of energy
sources (e.g., Wi-Fi, solar, thermal energy, etc.) which causes unpredictable
and frequent power outages~\cite{CampbellGD14,Rizzon:2013,Lui_2015,Jayakumar:2014,JayakumarRR14,lee2015powering,hglee14,wslee14,beeby2014,kellogg2016passive,aantjes2016testbed}. 

Energy harvesting systems use a small capacitor as an energy buffer and
intermittently compute only when enough energy is secured in the capacitor; when
it is depleted, the systems die. This is so-called {\it intermittent
computation}~\cite{lucia2017intermittent}.  With the intermittent nature in
mind, researchers adopt low-power microarchitecture with byte-addressable
nonvolatile memory (NVM) and offer a crash consistency mechanism to checkpoint
(backup) necessary data and restore them across power
outages~\cite{maeng2019supporting,WISP,WISPcam,MSP430FR5969,maeng2017alpaca,Clank,Chain,baghsorkhi2018automating,Chinchilla,dino,Ratchet}.

For the crash consistency, prior works including nonvolatile processors (NVP) approaches rely on  
voltage monitor based checkpoint schemes~\cite{
3us, Ambient,Lui_2015, ma2017incidental,
hibernus2015, balsamo2016hibernus++,jayakumar2014quickrecall}.
They checkpoint volatile registers---when the
voltage monitoring system detects the voltage drop below a defined threshold---by using
the buffered energy in the capacitor.  
In addition to the voltage monitor,
the schemes require non-trivial hardware modifications such as 
nonvolatile flip-flops, that must be laid out next to volatile flip-flops
for fast backup/restoration, special hardware checkpoint/controller logic, and 
additional capacitors for the voltage monitor.

Even worse, the voltage monitor may cause stability issues such as excessive
leakage or crack of the capacitors leading not only to reduced capacitance~\cite{capcrack,CapLeak}
but also to voltage detection delay with unexpected cold-start
glitch~\cite{su2017ferroelectric}. To mitigate the issues, 
 existing works aggressively increase the voltage threshold of the system wake-up/backup~\footnote{Without the
voltage monitor, the wake-up voltage can be set between
1$\sim$1.8V~\cite{3us,Ambient,msp430}, which is about 1.5$\sim$3X lower than
that of the
state-of-the-art work~\cite{su2017ferroelectric,wang2016storage}.}. Consequently,
they waste hard-won energy with making no forward progress until such a high voltage is secured to wake up the system for sure.

With that in mind, Hicks~\cite{Clank} proposes a voltage monitor free crash
consistency scheme called Clank by implementing idempotent
processing~\cite{Ratchet} in hardware. To identify idempotence
violation due to antidependence---i.e., WAR (writer-after-read)
dependence, Clank uses special dependence tracking hardware modules including
several memory buffers that keep track of every load/store
instruction.  Once such an antidependent load-store pair is detected, Clank
holds the store in the write-back buffer. If any of the buffers becomes full, Clank
checkpoints all registers, flushes the write-back buffer to nonvolatile
scratchpad, and copies the flushed data to NVM. 

Unfortunately, there are two major problems with Clank.  First, although 
nonvolatile scratchpad is used for performance reason, it is not clear how to
realize nonvolatile yet fast SRAM with current and future technologies.
One alternative is to exploit NVSRAM~\cite{xie2016emerging} (i.e., 3D stacking
of SRAM over nonvolatile memory) which copies SRAM data to the underlying slow
nonvolatile part right before power failure. However, this also requires the
voltage monitor and the checkpointing/controller logic, rendering Clank
vulnerable to the same voltage monitor issues.  Second, Clank can cause a
significant slowdown due to frequent overflows in the memory buffers; although 
using much larger buffers could mitigate the problem, it would increase the
dependence tracking cost in terms of the resulting power and delay.  According
to the Clank paper~\cite{Clank}, the performance overhead can be more than 20\%
even with the unrealistic assumption of the nonvolatile scratchpad. 

In summary, prior works suffer from the high hardware cost, low energy
efficiency, and run-time overhead. Thus, there is a compelling
need for a practical crash consistency solution that works for commodity
in-order processors without a significant slowdown.

To achieve low-cost yet performant intermittent computation, this paper
presents \name, a novel architecture/compiler co-design scheme that works for
commodity in-order processors used in energy harvesting systems.  To achieve
crash consistency without requiring unconventional architectural support,
\name leverages speculation assuming that power failure is not going to occur
and thus holds all committed stores in a store buffer (SB)---as if they were
speculative---in case of mispeculation.  \name compiler first partitions a
given program into a series of recoverable regions with the SB size in mind, so
that no region overflows the SB during the region execution.  When the program
control reaches the end of each region, the speculation turns out to be
successful; therefore, \name releases all the stores of the region, which have been
buffered in the SB, to NVM.

In case power failure occurs during the SB release, which may lead to incorrect
recovery, \name proposes a {\it two-phase release}, i.e., first draining the SB
entries to a proxy buffer allocated in NVM and then copying them to the primary
main memory area in NVM. That way, either the buffer or the primary main memory
can always be intact no matter when power is lost.  If power failure occurs
during the execution of a region, all its stores buffered in the SB disappear
because it is volatile.
The implication is that such mis-speculated stores---committed before power failure---cannot affect any program state in NVM at all.
Consequently, the interrupted region can be restarted with consistent program
states in the wake of power failure. 

While \name effectively provides crash consistency, the
region-based speculation window causes pipeline stalls at the end of each
region due to the two-phase SB release. Since it consists of NVM writes that
are the most time-consuming instruction, the stalls are rather long leading to
a significant performance overhead.  To hide the long NVM write latency of the
two-phase SB release, \name overlaps the SB release of the current region with
the speculative execution of the next region.  Such instruction level
parallelism (ILP) gives an illusion of out-of-order execution on top of the
in-order processor, achieving a speedup of more than 1.2X.

Finally, \name can take advantage of direct memory access (DMA), another
existing hardware component in commodity energy harvesting systems; a DMA
module has already been adopted in MSP430 series microcontrollers, and it
accelerates memory-to-memory copy in the FRAM-based nonvolatile
memory~\cite{MSP430FR5994, FRAMspeed}.  With the help of the DMA, \name can not
only accelerate the 2nd phase of the SB release, i.e., memory copy from the
proxy buffer entries to primary main memory locations, but also improve the ILP
efficiency, delivering significantly higher performance than all prior works!

Our contributions can be summarized as follows:
\begin{compactitem}
\item Unlike prior works, \name does not require any expensive hardware
modifications. \name's intelligent compiler-architecture interaction provides
commodity microarchitecture with crash consistency, achieving truly recoverable
intermittent computation at a low cost.

\item \name achieves high performance intermittent computation. The proposed ILP techniques allows \name to effectively hide the long latency of NVM writes in a program.
It turns out that the ILP is the {\bf main} reason for \name's high performance; \name
achieves speedups of 1.19X and 1.23X for non-DMA (i.e., ILP alone) and DMA
cases, respectively.  Overall, \name outperforms the state-of-the-art
nonvolatile processor by 11\% on average and up to 26\% when there is no power
outage.

\ignore{
\item \name achieves high performance intermittent computation by making an
illusion of out-of-order execution on top of an in-order processor. \name hides
the NVM write latency of the SB release, which is incurred at the end of each
recoverable region, by speculatively executing the next region. Such an ILP
optimization is the {\bf major} source of \name's performance gain; \name
achieves speedups of 1.19X and 1.23X for non-DMA (i.e., ILP alone) and DMA
cases, respectively.  Overall, \name outperforms the state-of-the-art
nonvolatile processor by 11\% on average and up to 26\% when there is no power
outage.
}

\item In the context of voltage-monitor-free energy harvesting systems, \name
can decrease the wake-up voltage by 1.5$\sim$3X compared to the
state-of-the-art
work~\cite{su2017ferroelectric,wang2016storage,lee2015powering}, which
leads to a much higher energy efficiency. The experimentation with frequent
power outages demonstrates that \name consumes 2$\sim$3X less energy 
than the state-of-the-art work.
\ignore{
and
achieves a speedup of 1.8$\sim$3X compared to the state-of-the-art work.
\todo{speedup does not match with that in in the prior bullet point}
}

\end{compactitem}


\section{Background and Motivation} \label{sec:background} 
\ignore{

This section introduces basic architectural concepts used throughout the
paper and points out the limitations of existing software/hardware approaches to
intermittent computation for energy harvesting systems.
}
\subsection{Architecture for Energy Harvesting}
\label{sec:model}
Given the power-hungry nature of energy harvesting systems, prior works have
opted for extremely low-power in-order processor architecture.  Despite the low
performance of the in-order processor, it is more suitable than power-consuming
and complex out-of-order processor architecture
~\footnote{Some prior works
propose to use out-of-order processors or even hybrid cores equipped with both
in-order and out-of-order pipelines. 
However, they assume strong energy harvesting source that can deliver 
stable power
for the out-of-order execution~\cite{ma2017spendthrift,ma2015architecture}.}.
Since power outages become the norm of program execution in energy harvesting
systems, their processors should have byte-addressable nonvolatile memory (NVM)
for efficient backup/recovery across the power outages.  For example, TI's
MSP430FR series of microcontrollers (MCU), which are popularly used in energy
harvesting systems, have already integrated FRAM as main memory for their
in-order processor core.

It is important to note that \name targets the commodity MCU made up of
FRAM-based nonvolatile main memory and conventional microarchitecture without
cache as prior
works~\cite{Ratchet,Clank,3us,Lui_2015,Xie_2015,choi2019achieving}.  In the
absence of cache, only data in a processor core, i.e., registers, are transient
and will be lost when power is cut off. Thus, registers need to be
checkpointed---i.e., saved in nonvolatile memory (NVM)---for their safe
restoration in the wake of power failure.

\subsection{Crash Consistency}
For crash consistency,
energy harvesting systems often makes a checkpoint to save the volatile
registers into the NVM during program execution and rollback to the checkpointed
states in the wake of power failure.  However, NVM alone cannot ensure correct
recovery due to memory inconsistency in which
data in NVM is corrupted across a power outage.  Suppose a simple NVM data
increment after checkpoint, i.e., `$checkpoint; i++$' where $i$ is
allocated in NVM and its initial value is $0$.  If the code is executed with no
power failure, the expected value of $i$ must be 1. However, if power is cut off
after the increment, the system rollbacks to the point where the checkpoint were
made and ends up re-executing the increment again. Since the $i$ has already
been updated before the power failure, the system leads to an unexpected output, i.e., 
$i = 2$. Thus, the memory inconsistency leads to incorrect recovery.
\ignore{
The
takeaway is that memory consistency is caused by the NVM store before power failure and its read when power comes back.  the wake of reading its data
}

\subsubsection{Software Based Approaches}
Prior work called Ratchet~\cite{Ratchet} identifies and eliminates the memory
inconsistency by using {\it idempotent processing}~\cite{de2012static,de2013idempotent,clover,16Tecs,bolt,ido2018,kim2020penny}. As the cause of the memory
inconsistency,  Ratchet recognizes the non-idempotent memory
access pairs comprised of a load and a subsequent store to the same memory
location, i.e., antidependence also known as Write-After-Read (WAR) dependence.
In the example of the NVM data increment `$i++$', the load and the subsequent
store to $i$ is such a non-idempotent memory access pair. To get around the
antidependence, the Ratchet~\cite{Ratchet} compiler automatically partitions an
input program to a series of idempotent regions so that each region has no
antidependence. That way, any interrupted idempotent region can be safely
re-executed without memory inconsistency when power comes back. Other prior
works ask users to partition a program into tasks and leave 
a log of data in each task---before they become inconsistent due to antidependence---along with checkpointing volatile registers that are inputs to some later tasks~\cite{maeng2017alpaca,Chain,dino}.
Unfortunately, all the software based crash consistency schemes often cause a significant
slowdown (50$\sim$400\% run-time overhead), which leads to the advent of hardware
based approaches.

\subsubsection{Hardware Based Approaches}
\label{sec:hardwareprior}
To reduce the overhead and eliminate additional burdens to end-users, prior
hardware approaches propose architectural support that relies on either (1) voltage monitoring system~\cite{3us, Ambient,
Lui_2015, ma2017incidental, hibernus2015,
balsamo2016hibernus++,jayakumar2014quickrecall} or (2) hardware-implemented idempotent processing~\cite{Clank}.

\paragraph{Voltage Monitoring Approaches}
For hardware approaches to crash consistent energy harvesting including nonvolatile processors~\cite{3us, Ambient, Lui_2015,
ma2017incidental}, a common requirement among many prior works~\cite{3us, Ambient,
Lui_2015, ma2017incidental, hibernus2015,
balsamo2016hibernus++,jayakumar2014quickrecall} is that they use a voltage monitor to
ensure the sufficient energy---left in the capacitor---enough for taking a full
checkpoint of volatile registers; the voltage monitor alerts the processor to take a checkpoint
when the monitor estimates that the capacitor contains the energy large 
enough to complete the register file checkpointing.
Similarly, the voltage monitor wakes up the processor when the sufficient energy is secured for successfully restoring the checkpointed values of the volatile registers. 
In this way, the voltage monitor systems can always make a checkpoint at the point right before power
failure and resume from exactly the same point when power comes back.
Since the systems never re-execute the code finished before power failure, they are
free from the memory inconsistency. 

However, this comes with an expensive hardware cost. In fact, the voltage monitor
based approaches require non-trivial hardware modifications such as
nonvolatile flip-flops, that must be laid out right next to volatile flip-flops
for the value of registers to move fast in between on their backup/restoration, special hardware checkpoint/controller logic, and the additional capacitors necessary for estimating the voltage level.

Even worse, the voltage monitor approaches have a precision issue.
In other words, it is very difficult to reliably estimate 
whether there is only sufficient energy left in the capacitor to
take a checkpoint without power failure. That is mainly because the estimation requires predicting the future rate of
charge-discharge. In fact, the capacitor can suffer from unstable discharge
due to in-field aging problems such as cracks~\cite{capcrack,CapLeak}. Increased discharge rate can make the voltage monitor underestimate the future
rate of charge dissipation. Unfortunately, this causes partial (unfinished) checkpointing, leading to incorrect recovery.
Even though the voltage monitor may address the unpredictable discharge rate problem
by pessimistically overestimating the discharge rate, this inevitably
increases the checkpointing frequency, thereby hurting the performance, energy efficiency,
and wear-out rate of the NVM.  
\ignore{
Moreover, the energy buffer is
vulnerable to physical access
attack~\cite{cronin2018collaborative,cronin2017danger}, whereby the attackers
can shorten the capacitor or detach it from the device then inject malicious
voltage fluctuation into the target board. If so, attackers can maliciously
cause the voltage monitor to create partial checkpoints that lead to malicious
memory inconsistency problems ~\cite{cronin2018collaborative,cronin2017danger}.
Due to their ubiquitous nature, energy-harvesting devices are particularly
susceptible to physical attacks. 
}

To address the capacitor issues and inaccurate timing error 
of the voltage monitor, 
more recent works
have developed precise and reliable voltage detection technologies~\cite{su2017ferroelectric}.
However, to ensure safe checkpoint/recovery, they increase the wake-up/backup voltage thresholds of energy harvesting systems by almost 2$\sim$3X
compared to the previous generation of voltage monitors~\cite{3us, Ambient};
the old voltage monitor approaches can start at about 1$\sim$1.8V~\cite{3us,
Ambient,msp430manual}, but the recent works set the threshold up high
as 2.7$\sim$3V~\cite{su2017ferroelectric,wang2016storage,lee2015powering}.
Consequently, they waste hard-won energy without making progress until the sufficient voltage is
provided to wake up the system.
Finally, they also found out that the
backup/recovery controller could make a wrong decision on a reboot time---which causes a potential checkpoint (data) corruption problem---due to the cold-start voltage spark or an unexpected glitch~\cite{su2017ferroelectric}.

\paragraph{Voltage Monitor Free Approaches}
To address the above issues, Clank~\cite{Clank}
proposes a voltage-monitor-free processor design by implementing the idempotent
processing in hardware. 
In detail, Clank
monitors all memory accesses (load/store) at run time with several memory
buffers such as write-back, read-first, write-first, and address prefix
buffers. By sweeping read-first and write-first buffers, Clank keeps track of
antidependent load-store pairs that make it impossible to perform idempotent
processing and thus lead to memory inconsistency in the wake of power failure.

In particular, once an antidependent store is detected, Clank holds it in the
write-back buffer; non-antidependent stores are directly merged into NVM.
If any of the buffers is about to overflow and unable to accommodate any further memory instruction (address), 
Clank alerts the processor to checkpoint all its registers,
flushes the write-back buffer to nonvolatile scratchpad emptying out
other buffers as well, and copies the flushed
data eventually to nonvolatile main memory.
Note that since it holds the antidependent store in the write-back buffer, Clank requires
every load to check the write-back buffer first in case of the store-to-load forwarding.

Unfortunately, Clank suffers from two significant problems that prohibit its
adoption.
First, although Clank takes advantage of nonvolatile scratchpad---much faster
than NVM---for performance reason, there is no current technology to realize
nonvolatile yet fast SRAM in reality.  Clank may leverage NVSRAM, a 3D stacking
based hybrid design of SRAM and NVM~\cite{xie2016emerging}, which copies SRAM
data to the slow nonvolatile part right before power failure. However, NVSRAM
also requires the voltage monitor and the necessary checkpointing/controller
logics, rendering Clank vulnerable to the same voltage monitor issues.  Second,
Clank may involve frequent checkpoints due to overflows in its memory buffers,
thus degrading the performance significantly. 

While Clank proposes to increase the size of the buffers for less overflows, it presents another---potentially more serious---problem in terms of the resulting hardware and energy costs.
To a large extent, enlarging the buffers puts significant pressure on the design of CAM (content addressable memory) structure for Clank's associative searches of the buffers; in fact, the size of load/store queues has scarcely increased at all in the last decade for the same reason.
Apart from the additional power consumption on the larger buffers, their wire delays might lead to significant energy consumption, possibly making Clank inappropriate for energy-harvesting systems.

With the reasonable size of the buffers, the performance overhead of Clank can be more than 20\%
even with the unrealistic assumption of the nonvolatile scratchpad~\cite{Clank}. 
With the deficiencies of all above prior works in mind, 
we seek to develop a practical crash consistency solution that works for commodity processors
without a significant run-time overhead.

\section{\name Approach} \label{sec:overview} \name is a low-cost architecture/compiler co-design scheme that enables reliable crash
consistency \ignore{on top of commodity energy harvesting systems (in-order
core)}without significant energy and performance overheads.
%
This section first presents the basic design of \name: (1) hardware design, (2) compiler
support, and (3) architecture/compiler co-design. The optimization techniques of \name are deferred to
Section~\ref{sec:impl}.

\subsection{\name Hardware Design}
\label{sec:hwdesign}
The design philosophy of \name is to leave the commodity microcontroller (MCU)
architecture~\cite{mehta2016wearcore, minorCPU, MSP430FR5994}---used in
energy harvesting systems---almost as is and enable high performance intermittent
computation without expensive hardware modifications.
\ignore{
On top of the current commodity TI's MSP430FR series of MCUs~\cite{MSP430FR5994},
\name only leverages the store buffer and DMA.
For optimization, \name also leverages ISA support,
which will be discussed in Section~\ref{sec:alias}.
}

\ignore{Figure~\ref{fig:new_model} describes the architecture diagram of \name.
\begin{figure}[!h]
\centering
\centerline{\includegraphics[width=\columnwidth, angle=0]{fig/design.pdf}}
\centering
\caption{\name Hardware micro-architectural model}
\label{fig:new_model}
\end{figure}}





\ignore{
In short, \name leverages {\it SB} and DMA accelerator 
to address memory consistency, non-deterministic
path issue, and even improve
performance.
This paper discusses about two key components in next sections.
}

\subsubsection{Store Buffer for Power Failure Speculation}
\label{sec:storebuffer}
Store buffer has been adopted for other commodity in-order
MCUs~\cite{mehta2016wearcore}, e.g.,
ARMv8-A core implementations~\cite{minorCPU}, mainly to handle mispeculation 
such as branch misprediction~\footnote{Currently, the processors used in energy-harvesting systems have no branch predictor as in MSP430 MCUs, NVPs (nonvolatile processors), Clank, and \name.}.
To achieve lightweight crash consistency, \name proposes to exploit such a
 store buffer (SB) for a different type of speculation.



The difference is that \name uses a region-level speculation window,
guessing whether each region is likely to finish without interruption due to power
failure. In other words, \name leverages the store buffer (SB) to hold committed
stores of each recoverable code region during its execution---since they are treated as speculative---until the
program control reaches the end of the region (i.e., the region boundary) where 
the speculation turns out to be successful and thus all the buffered stores are
released.

Note that this speculation approach never allows the stores of any regions
being interrupted by power failure to be written to primary main memory (NVM).
If power failure occurs, all buffered stores in the SB disappear
because it is volatile.  It is therefore impossible for the mis-speculated stores to affect
NVM. Consequently, the interrupted region can be restarted with consistent program
states in the wake of power failure. 
The takeaway is that speculative stores cannot be released to NVM until they
become non-speculative, i.e., their region finishes without power failure. 
As a result, \name can completely eliminate memory
inconsistency without adding multiple non-trivial microarchitectural components
required by nonvolatile processors and Clank.

It is important to note that \name splits the store buffer into two parts to
enable instruction level parallelism as will be shown in Section~\ref{sec:ilp}.
During the program execution, any two consecutive recoverable code regions exclusively occupy
one of the two parts in the SB. 
That is, each statically partitioned code region commits its stores to a
different part of the {\it SB} at run time. 
When the program control reaches each region boundary, \name drains to NVM only the stores in the part of the SB which is used by the region being finished.

As the major challenge in achieving correct crash consistency, \name should
maintain failure atomicity of the {\it SB} draining; otherwise, any partial draining
may result in the memory inconsistency
problem~\cite{Ratchet,Clank,su2017ferroelectric}. \ignore{ (see
Section~\ref{sec:overview}).} To overcome the challenge, \name leverages a 2-phase
SB release mechanism; \name first drains the committed stores from the {\it SB} to
a proxy buffer in NVM, and then copies the drained results from the buffer
to the primary main memory area in NVM. With the help of the 2-phase SB
release, either the buffer or the main memory can always remain intact no
matter when power is cut off.  This will be discussed in Section for more
details~\ref{subsub:checkpoint}.

\subsection{\name Compiler}
\label{sec:compiler}
\ignore{
and lessen the hardware modification cost~\footnote{
\name can eliminate extra hardware components such as counters and its control logic~\cite{TCCP} by compiler support.}.
}
To partition the program into such regions, \name compiler first counts the
number of stores while traversing the control flow graph (CFG)
of the program.  When the number of stores hits a threshold, i.e., a half the SB size, \name compiler cuts the current basic block---where the last store is counted---by placing a region boundary. Then, the
compiler analyzes the live-out registers of the resulting region and inserts a
checkpoint instruction to save them into a designated 
register file (RF) checkpoint storage in NVM. In particular, a PC register is saved at the end of each
region---which serves as a recovery point in the wake of power failure---so
that the forthcoming power failure will be recovered by restarting the next
region. 

Note that all inserted checkpoint instructions are normal store
instructions. Thus, according to \name's power failure speculation, they are first committed to the SB and then drained to NVM provided the speculation turns out to be successful.
Indeed, the region formation is a tricky problem due to the circular
dependence during the partitioning process. That is, the live-out register
checkpointing essentially adds store instructions to a region, and of course
the number of (added) stores determines the region boundary, which in turn
affects the live-out registers provided the region boundary changes. 

\subsubsection{Region Formation}
\label{sec:region}
To conduct the region formation, we leverage the algorithm used in our own prior work~\cite{16Micro}. In the following, we describe the high-level idea; for more details, readers are referred to the work~\cite{16Micro}.

\name first partitions an input program into common program structures such as
calls and loops. For this purpose, \name
places a region boundary at all the entry and exit points of functions.
Likewise, a boundary is placed at the beginning of each loop header. Next,
\name identifies the basic block that has region boundaries in the middle of
it, and splits it into separate basic blocks.  This allows the region
boundaries to always start at the beginning of basic blocks, which helps the next
step to compute the initial checkpoint instructions.  After finishing the
initial region formation, \name analyzes the regions to place live-out register
checkpoints (i.e., store instructions).

Then, \name compiler traverses the CFG in a topological order trying to combine
those initial regions into larger regions as much as possible. The region
combining can eliminate many checkpoints because the live-out registers of
preceding region(s) are often no longer live after being combined with
following regions.  During the traversal of each control flow path, \name
updates the sum of current and incoming basic blocks' stores from the beginning
of the latest region boundary along the path.  If the sum becomes greater than
a half of SB size (threshold) before the next region boundary is reached, \name
places a boundary to cut the region.  After that, \name compiler analyzes the
re-partitioned regions again to insert live-out checkpoints and possibly
repeats the re-partitioning process as long as there is a region that has more
stores than the threshold. 

In this way, it is guaranteed that the each partitioned region has at most as
many stores as a half of the SB size, i.e., the threshold. It would be a
mistake to take this to mean that all regions have exactly the threshold number
of stores; rather many regions could have less stores than the threshold due to
the re-partitioning process.

\paragraph{I/O Operations}
To the best of our knowledge, to support non-recoverable operation such as
I/O operation has remained as the open problem. 
That being said, since \name compiler
places a region boundary at function calls, the function that implements I/O
operations is treated as a separate region---though it cannot be recovered due to the I/O operation. 
We believe that \name can deal with I/O operations by simply checkpointing
necessary status---just before each I/O operation starts---so that the interrupted
I/O operation can be restarted in the wake of power failure.

\ignore{
To solve the problem, 
\name compiler first finds all
of basic blocks, and then places region boundaries at all of function entries
and exits, loop headers, memory fences, and atomic operations. 
Then, in such an initial region, \name inserts a checkpoint instruction, which
is an essentially store instruction, right after the last update of a register
that is used in later code to save the live-out register in NVM.
Then, the compiler \name compiler updates the number of stores for the current region with the sum of
original stores and the checkpoint stores. 
Once this initial process is done, \name checks each
basic block in a topological order. In between region boundaries, it
counts the total number of stores. 
If the number of stores exceeds the store buffer size, \name compiler cuts the
region, and re-calculate the live-out register checkpointing stores, and updates
the total number of stores.
The above process keeps repeated as long as the total number is greater than the
threshold (i.e., half size of SB).

The takeaway is that the \name compiler ensures the number of stores during the execution
of each region will not be greater than the threshold (i.e., half size
of SB) even after the checkpointing stores are inserted. 
}

\begin{figure*}[!htb]
\centering
\begin{minipage}[b]{0.49\textwidth}
\centering
\includegraphics[width=\columnwidth, angle=0]{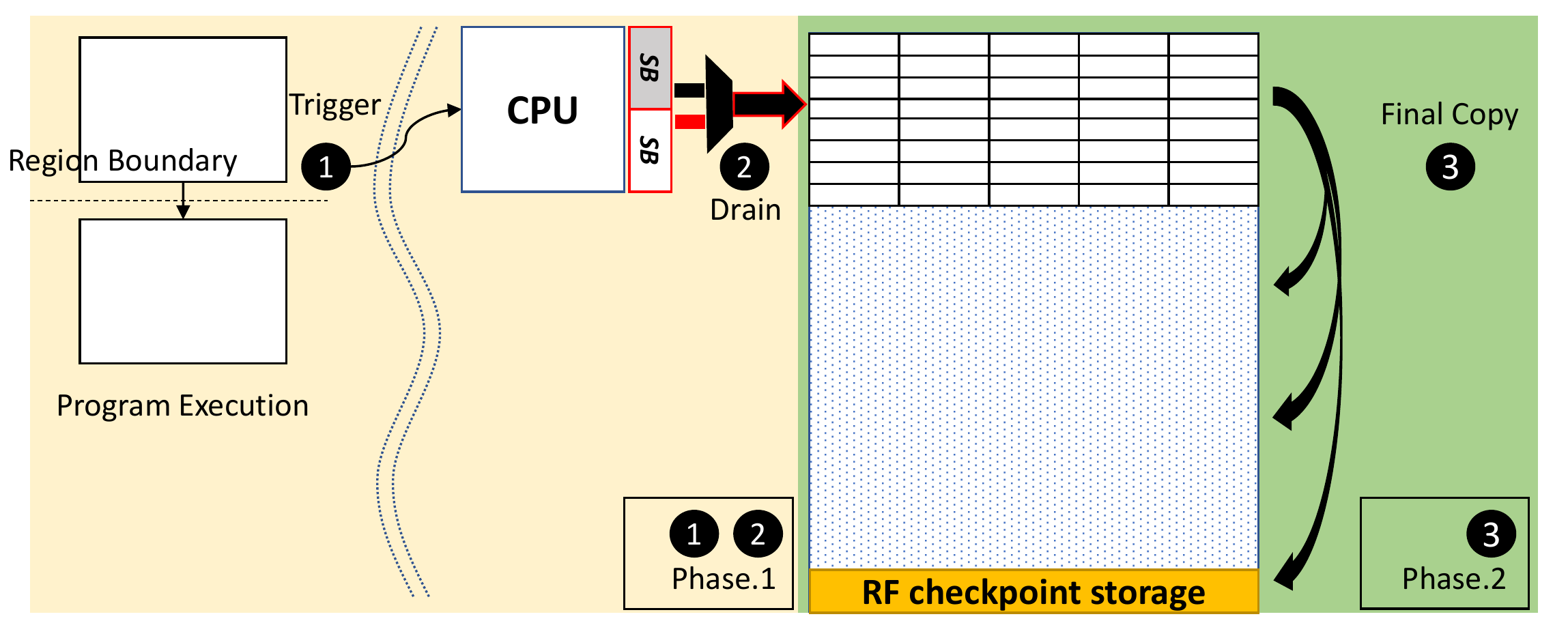}{\ (a)\ Region-based Checkpoint (normal case)}
\end{minipage}
\begin{minipage}[b]{0.49\textwidth}
\centering
\includegraphics[width=\columnwidth, angle=0]{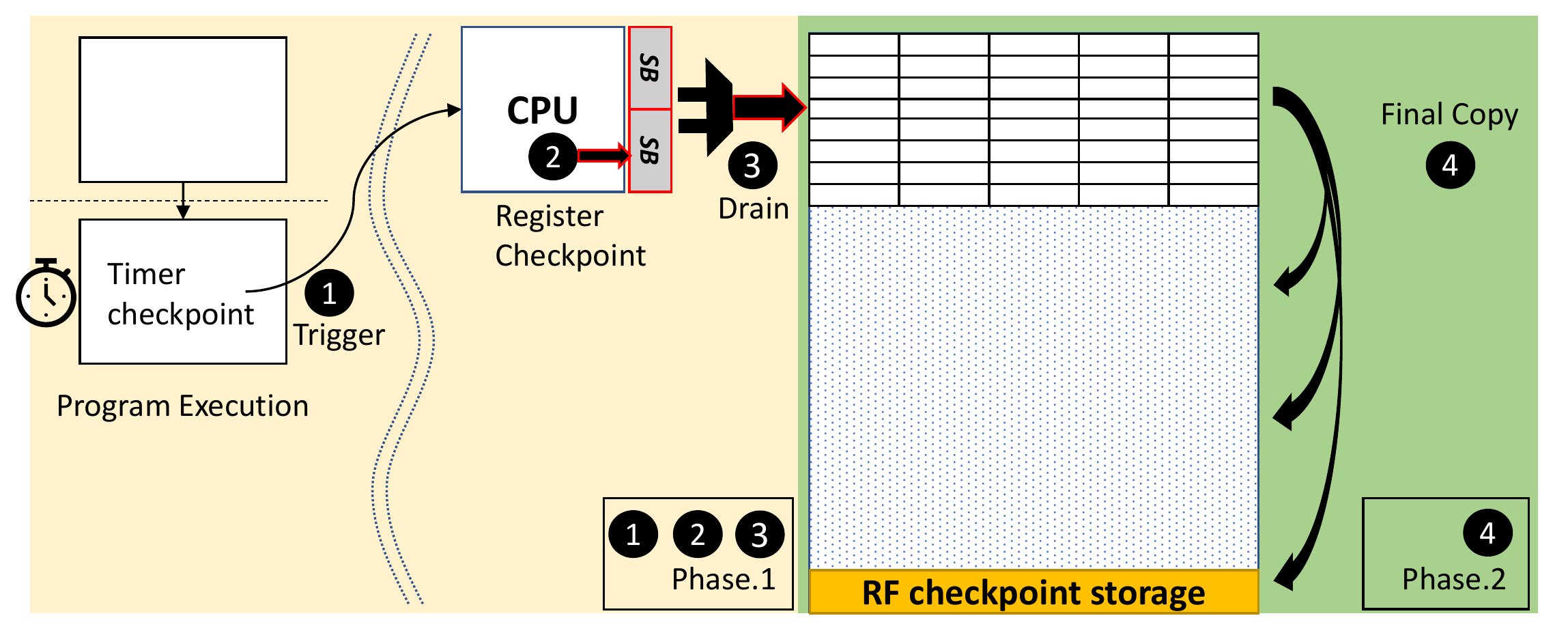}{\ (b)\ Timer-based Checkpoint (exceptional case)}
\end{minipage}
\caption{\name's checkpoint protocol for a normal case (a) and an exceptional case such as stagnation (b)}
\label{fig:checkpoint}
\end{figure*}

\subsection{Architecture/Compiler Co-design}
\subsubsection{2-Phase Store Buffer Release Protocol}
\label{subsub:checkpoint}
To achieve failure-atomic store buffer (SB) release, which is required for safe power
failure recovery without memory inconsistency, \name drains SB to NVM using a
2-phase mechanism.  When each region is ended, i.e., program control
reaches the end of each region boundary, \name first drains the committed
stores to a proxy buffer allocated in NVM and in turn moves the drained data
from the buffer to the primary main memory in NVM.  Figure~\ref{fig:checkpoint}
describes how the 2-phase SB release protocol works for (a) a normal case
(region based) and (b) an exceptional (watchdog timer based) case.  First, in
the normal checkpoint case, the system  (\ding{182}) triggers the SB release
when each region boundary is reached during the program execution.  Then, \name (\ding{183}) drains one
part of {\it SB}---which corresponds to the region being finished---to the
proxy buffer in NVM. As soon as the draining is completed, \name (\ding{184})
copies all the buffered data to primary main memory locations.

Second, \name also supports an exceptional (watchdog timer based) case. 
In particular, if a compiler-partitioned region is excessively long, the system
might be unable to make forward execution progress because of re-executing the
interrupted region again and again across power outages; this paper calls this
live lock like situation {\it stagnation}.
To avoid the stagnation, \name dynamically checkpoints registers to SB at the
expiration of a watchdog timer---which can be adjusted at run time taking into
account the dynamic power failure behaviors as will be shown in
Section~\ref{sec:forward}.  Figure~\ref{fig:checkpoint}(b) shows how the
dynamic checkpointing works with the 2-phase SB release.  When the watchdog
timer (\ding{182}) expires, \name (\ding{183}) immediately
checkpoints (stores) all registers and commits them to the \textbf{idle part of
SB}--not used by the current region. \name (\ding{184}) then drains full {\it
SB} to the proxy buffer in NVM.  
When two parts of {\it SB} are completely drained, \name (\ding{185}) makes the buffered data
moved to the primary NVM locations in the same way as a normal case; note that,
the watchdog timer is disabled during the 2-phase SB release process.

The 2-phase SB release mechanism protects both proxy buffer and the primary data
in NVM---by managing a check bit for each---against the partial SB draining that may fail to recover from power failure.
The first bit {\it isDrain} is devised for `Phase 1' release, and it is set when the part of
{\it SB}, which corresponds to the region being ended, is completely drained into the proxy
buffer in NVM; in the exceptional case shown in Figure~\ref{fig:checkpoint}(b), the bit is set when both parts of SB
are drained completely. The second bit {\it isComplete}---devised for `Phase 2'
release---is set when all the data in the proxy buffer are completely moved to
primary main memory locations in NVM.  These two check bits help \name to restore
correct data in the wake of power failure, and the next section discusses more
details about the
recovery protocol.

\subsubsection{Recovery Protocol}
\label{sec:recovery}
\name provides a safe recovery protocol to address potential
memory inconsistency problem across power failures.
There are three possible cases of power failure that differ in terms of their
failure point in the timeline.
First, a power failure can occur during {\it SB} draining (Phase 1 release),
i.e., {\it isDrain} bit is not set.  In
this case, \name simply ignores the SB data drained to the proxy buffer in NVM; the SB
contents all disappear due to the volatility of the SB.
To resume the interrupted region in the wake of the power failure, 
\name first restores the saved register values including the recovery PC from the RF checkpoint storage in
NVM and jumps to the PC. Note that it points to the beginning of the
interrupted region at the moment.  Although at the end of the region, a compiler-inserted checkpoint successfully
saved a new recovery PC that points to the
beginning of the next region, it was not written to neither the proxy buffer
nor the RF checkpoint storage in NVM because of the power failure occurred during the 'Phase 1' release.

Second, a power failure can occur during the copy from the proxy buffer to the
primary main memory in NVM (Phase 2 release).
In this case, since the {\it isDrain} bit has been set, i.e., the recovery PC
checkpoint at the end of the current region was successfully written to the proxy
buffer, \name does not rollback to the beginning of the current region. Instead,
\name does redo the Phase 2 release, i.e., moving the proxy data to the primary main
memory in NVM.  Then, as usual, \name restores the saved register values
including the recovery PC from the RF checkpoint storage in
NVM and jumps to the PC for recovery.

Third, a power failure can occur outside of the 2-phase release, i.e., in the
middle of a region, \name recognizes such a case by checking the both bits,
i.e., {\it isDrain} and {\it isComplete} are set. Here, the recovery process is
simpler compared to the above two cases.  \name just restores the saved
register values including the recovery PC from the RF checkpoint storage in NVM
and jumps to the PC that should point to the beginning of the region
interrupted by the power failure. The takeaway is that according to the status
of the two check bits, \name takes appropriate actions for correct recovery,
thereby ensuring truly-recoverable intermittent computation no matter when
power is lost and how often it occurs~\footnote{The recovery protocol can be further optimized by using only one bit. For safe recovery, the check bit is set to 1 when the `Phase 1 release' is finished, and it is reset to 0 when the `Phase 2 release' is finished. This implies that the bit is always zero when a new region starts. In the wake of power failure, if the bit is 0, \name simply restarts the interrupted region by restoring registers and jumping to the recovery PC; otherwise, \name first redoes the 'Phase 2 release' and then restarts the region as usual.}.

\section{Optimization} \label{sec:impl} \begin{figure*}[!htb]
\begin{minipage}[b]{0.49\textwidth}
\centering
\includegraphics[width=\columnwidth, angle=0]{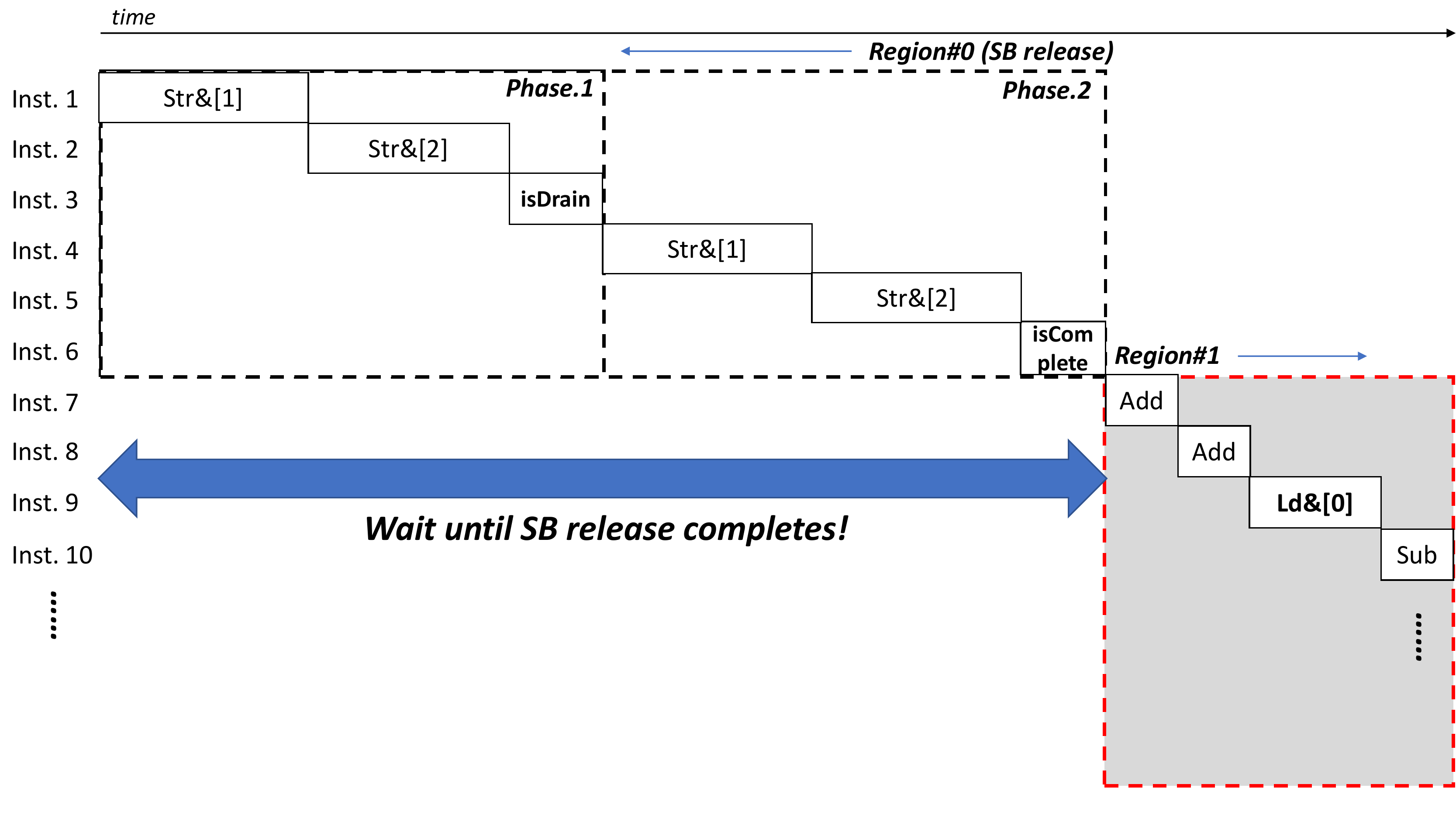}{\ (a)\ \name without ILP}
\end{minipage}
\begin{minipage}[b]{0.49\textwidth}
\centering
\includegraphics[width=\columnwidth,  angle=0]{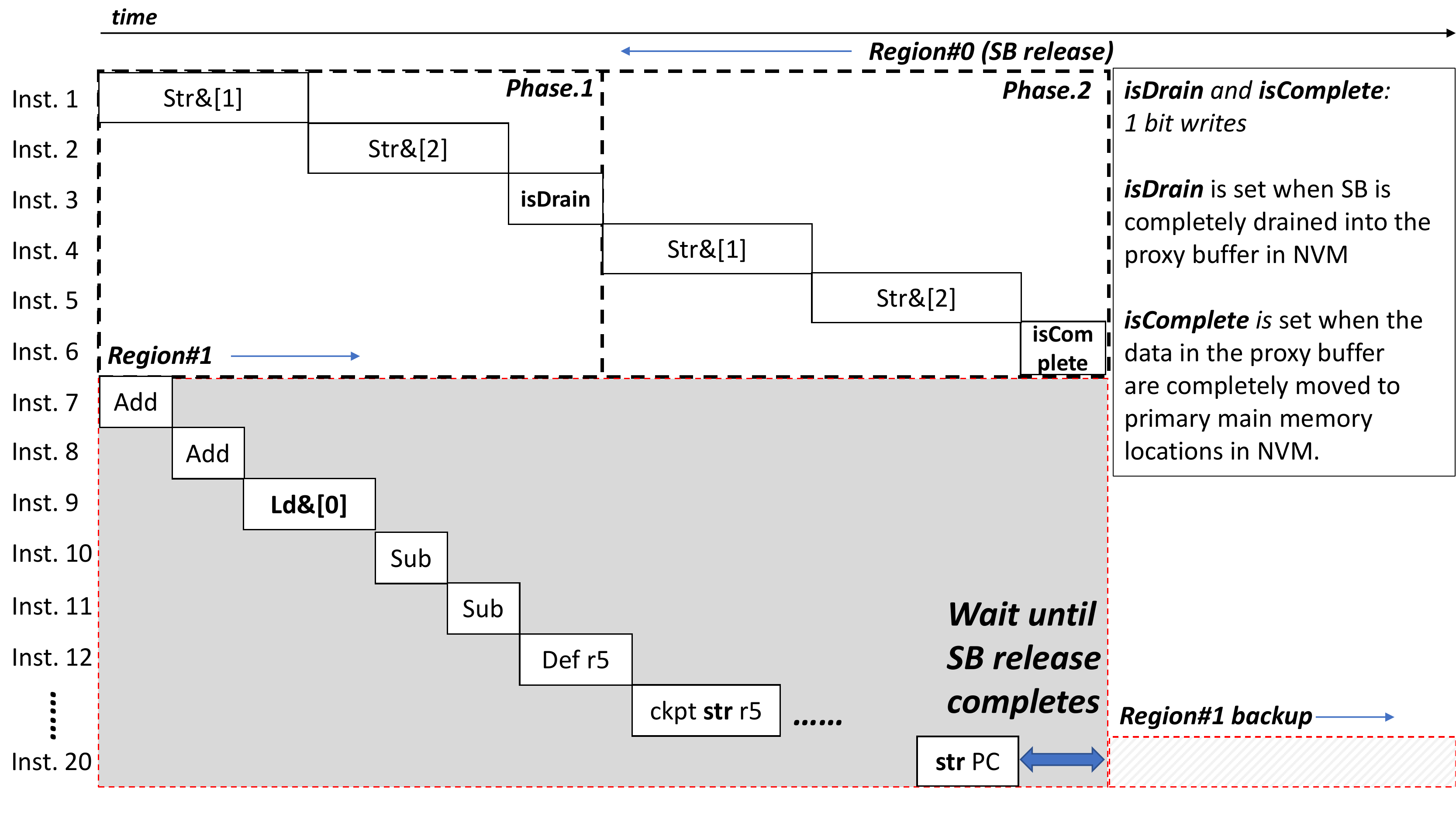}{\ (b)\ \name with ILP}
\end{minipage}
\caption{Performance benefit thanks to ILP. DMA is not enabled here, though it can accelerate the 2nd phase of the SB release.} 
\label{fig:ILP} 
\vspace{-4mm}
\end{figure*}

To avoid potential memory inconsistency, the 2-phase SB release mechanism 
requires double persistent writes for all stores. Unfortunately, this incurs
significant performance overhead consuming hard-won energy---for such expensive NVM writes---that would otherwise could be used for making further forward execution progress.
To address the overhead problem, \name optimizes the 2-phase SB release
by enabling instruction level parallelism (ILP). That is, \name does not wait until its 2-level
SB release is finished; rather it speculatively executes the next region's
instructions while the SB release is pending.
This section describes the implementation details of such an optimization: (1) how to reliably
enable the ILP execution on an in-order processor without memory inconsistency and (2) how to adapt the ILP for the intermittent computation where the frequency of power outages varies.

\subsection{Instruction Level Parallelism}
\subsubsection{Enabling ILP execution}
\label{sec:ilp}
\name enables instruction level parallelism to hide long NVM write latency by
overlapping them with the next code region execution~\footnote{
Similarly, TSO\_ATOMICTY~\cite{wang2013tso_atomicity}
leverages the overlapped region execution for atomic-region based dynamic optimizations. 
However, that is devised for multi-core out-of-order processors to achieve more thread interleaving. Also the store queue design and the region formation algorithm are different from those of \name.}.  
Figure~\ref{fig:ILP} shows how the instruction level parallelism (ILP) works when two consecutive regions (i.e., Region\#0 and Region\#1) are executed. 
Figure~\ref{fig:ILP}(a) describes a non-ILP case; when {\it
SB} starts its draining to NVM using the 2-phase release mechanism, the system needs to wait until the both phases finish to ensure the SB data is safely written to the primary memory. Since partial SB release can cause memory inconsistency, the power failure recovery might fail.
Figure~\ref{fig:ILP}(b) shows how \name hides such a long latency of NVM
writes. By overlapping the NVM writes during entire 2-phase SB release with the speculative execution of the next code region, \name is
able to execute more instructions within a given time; as shown in
Figure~\ref{fig:ILP}(b), the ILP approaches executes 13 more instructions than the non-ILP approach that encounters stalls at the instruction \#6 due to the 2-phase SB release.

Note that once the speculative execution of Region\#1 is completed, \name should wait until the
SB release of the Region\#0 is finished rather than executing the next region
(Region\#2 not shown in the figure). This is necessary for achieving correct crash consistency.
The next section shows how \name solve this problem.

\subsubsection{Achieving ILP without Breaking Correctness}
\label{sec:hazard}
There are a few challenges \name must overcome to achieve the ILP optimization
for correct recovery.
First, \name should avoid inserting a store to the 
{\it SB} during its draining; otherwise, it may incur the data hazard or race condition on the store buffer.
To address this challenge, \name lets each code region alternatively
use a different part of {\it SB}. Recall that \name splits the SB to two parts
for exclusive use of any two neighboring code regions.
For example, if a current region
inserts its stores to one part of {\it SB}, then the next region inserts its
stores to the other part of {\it SB}. That is, any two consecutive regions
exclusively use a different part of {\it SB} all the time.
However, it is still possible to insert a store to the same part of {\it SB}. 
For example, if the speculative region execution finishes too fast even before the 2-phase SB release of the previous region is completed, then executing the following region may overwrite data in the part of SB which is pending (being drained) for its
2-phase release.  To avoid this problem, \name conservatively waits 
at the end of the speculative region while the previous region's SB release is pending.

Second, load instructions should read the up-to-date data for correct
execution.  Suppose that a current load instruction needs to read data, but the
required data is placed in the part of SB which is being drained.  In this
case, the load instruction should be stalled for correctness purpose. To avoid such a delay, both parts of SB must be available for correct execution. With that in mind, \name does not invalidate the {\it SB} entries being drained until the program control reaches the end of the speculative region, i.e., the one following the prior region whose 2-phase SB release is pending.
That way, the load of the speculative region can 
read any written data of the prior region from its part of the SB---which is
being drained---without any stall.
Of course, when the load in a region is to read the data written by the same region,
its load can be served as usual using the conventional store-to-load forwarding
through its own SB.

\paragraph{Discussion}
One might argue that adding a SB in a simple in-order pipeline could reduce the core clock frequency as with modern processors where their SB must provide a dependent load with data within
L1 hit time to avoid complicating their scheduling logic.  However, we believe that \name is
free from this concern thanks to its architecture characteristics. Apart from
the use of in-order pipeline and low clock frequency ($\sim$25MHz) in energy harvesting systems, \name does not have a
cache (Section~\ref{sec:model}). The implication is that the SB
search has only to finish within NVM (i.e., FRAM) access time. Note that this
is always doable because each SB entry access is orders-of-magnitude faster
than FRAM access latency.  Consequently, \name causes neither clock frequency
reduction nor scheduling logic complication~\footnote{Technically, accessing 40
SB entries takes less than 1 cycle~\cite{gunadi2007position,gu2016nvpsim}}.  In
addition, \name can bypass SB searches for the majority of following loads;
Section~\ref{sec:alias} details the SB bypassing and necessary compiler
analysis.

\ignore{
how to reduce the search cost by eliminating unnecessary SB searches, and how to efficiently
modify the store buffer search logic in 

consumes a bit of energy due to the CAM search~\cite{gunadi2007position}.
}

\subsection{Stagnation-Free Intermittent Computation}
\label{sec:forward}
\name should address the \textbf{stagnation} problem
(Section~\ref{subsub:checkpoint}), which would otherwise waste the harvesting
energy in vain without making forward execution progress.  To ensure the
forward progress in the presence of frequent power outages, \name proposes
adaptive execution techniques~\cite{choi2019achieving,jung2005adaptive,adp10} that take into account dynamic power failure
behaviors.

The use of ILP optimization and the region-level speculation window may
increase power consumption compared to non-modified design, possibly causing
more power failures during intermittent computation. In light of this, \name
adaptively turns on/off the ILP and adjusts the speculation window according to
the power failure patterns in a reactive manner.

When the system suffers from power failures, \name first turns off the ILP
execution. Then, if the power failure happens in the same region more than
twice, which might be a sign of stagnation, \name turns on the watchdog timer
checkpoint.  Once the timer is expired, \name checkpoints registers to the
store buffer (SB) and performs the 2-phase SB release as shown in
Figure~\ref{fig:checkpoint}(b). Since the timer is set for it to be expired in
the middle of the stagnating region, \name can resume from the timer expiration
point in the wake of power failure---rather than jumping back to the beginning
of such a long region.  If the region still encounter another power failure,
\name decreases the watchdog timer to a half of the previous value. This in
effect doubles the frequency of the register checkpointing (and the 2-phase SB
release) and can be repeated to get out of any long stagnating region across
power outages.

On the other hand, if the system continues to make progress without a power
outage in which case \name assumes the system is under a good energy harvesting
condition, then it enables ILP and disables the watchdog timer approach.  With
this simple adaptive execution heuristic, \name can address the stagnation
problem and improve the performance by spending more harvested energy for
forward execution progress rather than wasting it for the re-executions of
stagnating regions.

\subsection{Energy-Efficient Store Buffer Search}
\label{sec:alias} In case of store-to-load forwarding, every load should
consult the store buffer (SB). However, this involves expensive CAM (content
addressable memory) based associative search in the SB. To address this issue,
\name (1) bypasses unnecessary SB searches and (2) designs a cost effective
SB search logic.

First, \name compiler statically checks if each load can be may- or
must-aliased to stores in the current and previous regions by leveraging alias
analysis ~\cite{andersen1994program, lattner04llvm,
vedula2018nachos,sui2016svf}. When no alias is found, \name compiler marks the
load instruction so that it can bypass the SB.  During the program execution,
if the processor detects such a special load instruction, it avoids the SB
search and directly accesses to primary main memory.

\begin{figure}[!htb]
\centering
\centerline{\includegraphics[width=\columnwidth, angle=0]{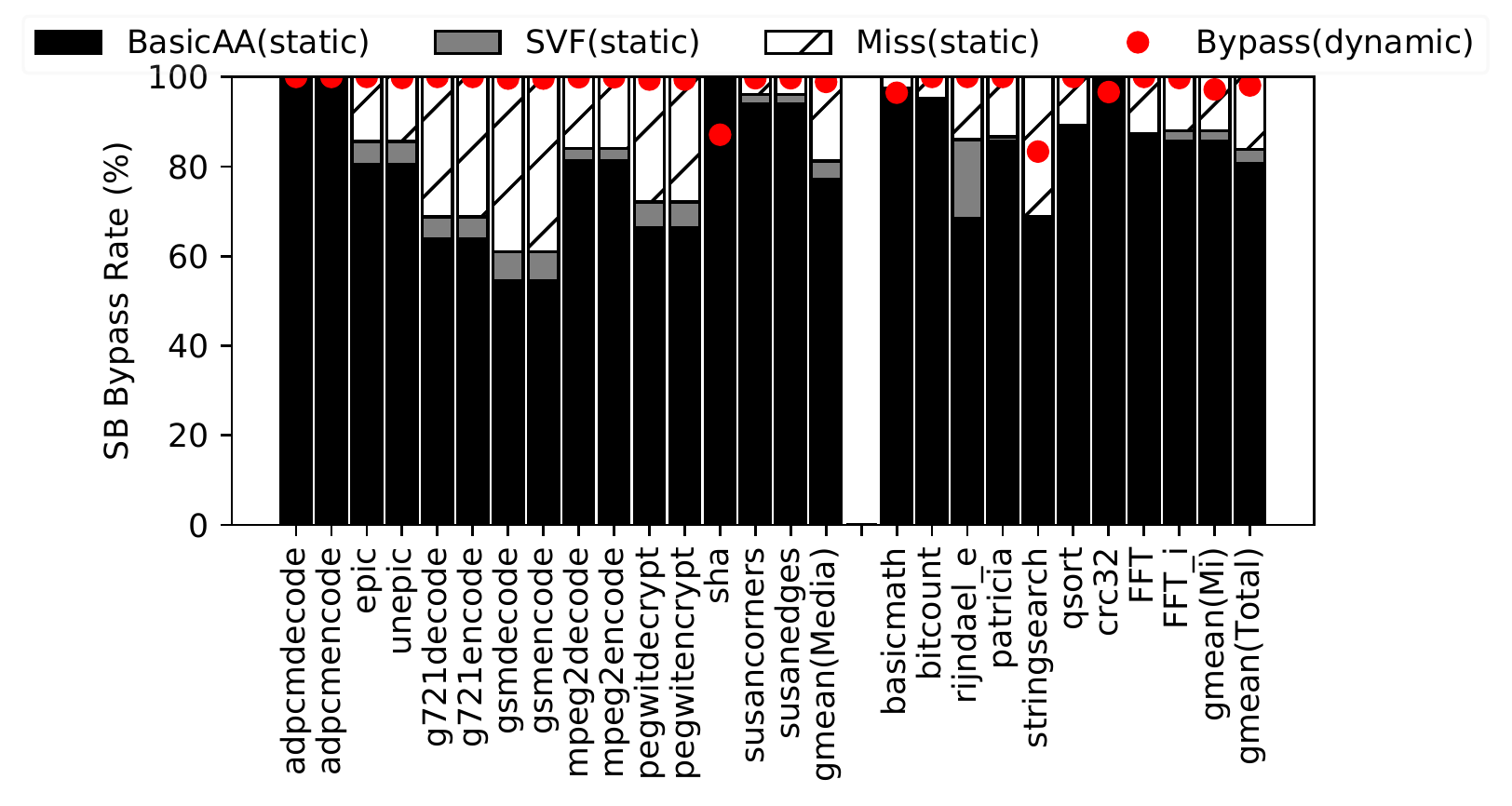}}
\centering
\caption{Store buffer bypass rates at compile time and run time. Both BasicAA 
and SVF are static alias analysis.}
\label{fig:aa}
\end{figure}

To see the impact of this compiler-directed SB bypass scheme, we conducted
measured how many load instructions could avoid SB searches at both compile
time and run time.  The experimental result demonstrates that a significant
number of loads is able to bypass the SB search. As shown in
Figure~\ref{fig:aa}, at compile time, more than 80\% of total load instructions
can be marked to bypass the SB search on average by using both basic alias
analysis (BasicAA)~\cite{lattner04llvm} and advanced alias analysis called SVF
(static value-flow analysis~\cite{sui2016svf}~\footnote{\name compiler could run
SVF---which is field- and flow-sensitive---successfully on top of program's
region-based control flow sub-graph; while such an advanced analysis is very
expensive for whole program analysis, our region-based (per-region) analysis
makes it possible to run the SVF for all the benchmarks we tested.}). At run
time, 98$\sim$99\% of dynamic loads turn out to be from the SB search. That is
mainly because many non-aliased loads are found in hot loops whereas aliased
loads are not. 


In particular, the promising results of high SB bypass rates motivate the different
design of the SB search mechanism. In other words, \name can afford a sequential search
logic rather than the expensive CAM-based associative search. This gives a freedom to use the SB for energy harvesting
system without worrying about the high power consumption required for the CAM search.

\begin{figure}[!h]
\centering
\centerline{\includegraphics[width=\columnwidth, angle=0]{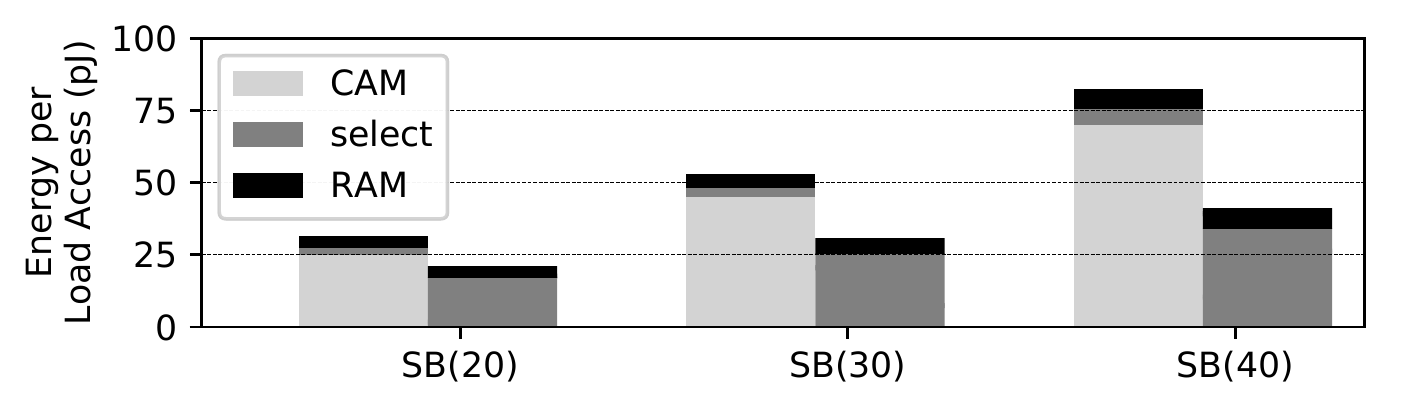}}
\centering
\caption{Energy consumption breakdown of different SB search schemes.
For each SB configuration on x-axis, the first and second bars represent
conventional CAM search and \name's sequential search, respectively.}
\label{fig:CAMenergy}
\end{figure}

To this end, we estimated the energy consumption and performance of both conventional CAM-based associative search
and sequential search by using CACTI~\cite{shivakumar2001cacti} with 90nm
technology~\cite{msp430manual} in the same way as prior work~\cite{gunadi2007position}.  
While the
conventional associative SB search is comprised of three components, i.e., CAM, select logic, and
RAM (buffer), \name can remove the CAM part thanks to the sequential search.
Figure~\ref{fig:CAMenergy} describes the energy
consumption breakdowns of the CAM-based associative search and the sequential search. 
When the SB is 40, the energy consumption of the associative SB search is about 2X
greater than the sequential search.

\begin{figure}[!h]
\centering
\centerline{\includegraphics[width=\columnwidth, angle=0]{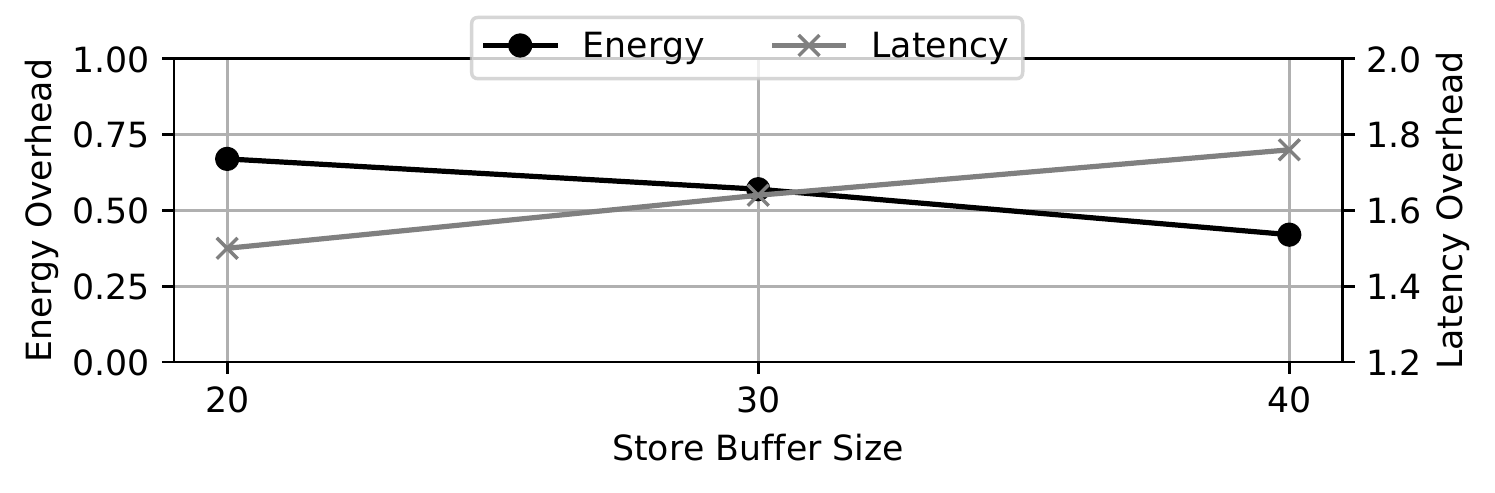}}
\centering
\caption{Normalized energy/latency overheads of the sequential SB search
compared to the CAM based associative search}
\label{fig:gsbX}
\end{figure}


We also
analyzed the latency overhead of \name's sequential search compared to the CAM search.
Figure~\ref{fig:gsbX} shows both the normalized access latency and the energy
consumption overhead.  The latency overhead is about 1.5$\sim$1.8X when the store
buffer size is 20$\sim$40, while the energy consumption reduction is 40$\sim$70\%.
Section~\ref{sec:eval} evaluates the impact of the both SB search schemes for
various benchmark applications.

\subsection{Direct Memory Access (DMA)}
\label{sub:dma}
Although ILP execution can hide the long latency of the 2-phase SB release, it does not
reduce the latency.  To accelerate the SB release, \name can opt for DMA
processing available in commodity energy harvesting microcontrollers (MCUs),
e.g., MSP430 series.  In fact, the DMA engine of MSP430
MCUs~\cite{MSP430FR5994} can speed up NVM data transfer, i.e., memory-to-memory
copy, by $\approx$4X faster then normal read-write based copy~\cite{FRAMspeed}.
In light of this, \name can use the DMA to accelerate the second phase (i.e.,
Phase 2 data copy shown in Figure~\ref{fig:checkpoint})  of the 2-phase SB
release.

However, care must be taken to perform the DMA processing because every data in
the proxy NVM buffer needs to be copied to the corresponding primary main
memory locations in a precise manner.  Currently, \name uses a single DMA
channel multiple times in a row. That is, the number of DMA operations is the
same as the number of the proxy buffer entries to be copied.  Although a series
of DMA copies seem to be not optimized, the DMA processing is still helpful
thanks to its 4X faster NVM copy.  It is important to note that due to the DMA
processing can improve the ILP efficiency as will be shown in
Section~\ref{sec:sensitivity}. That is because the prevention of the SB race condition lets the ILP mechanism conservatively wait at the end of the speculative region for the previous region to complete its 2-phase release (See Section~\ref{sec:hazard}).


\section{Evaluation} \label{sec:Eval} \label{sec:eval}
We implemented \name compiler techniques described in
Section~\ref{sec:compiler} using the LLVM compiler
infrastructure~\cite{lattner04llvm}. All the experiments were performed on the
gem5 simulator~\cite{Binkert11gem5} with ARM ISA,
modeling a single core 3-stage in-order pipeline as in
NVP simulator~\cite{gu2016nvpsim}. We compared \name to nonvolatile processor
(NVP)~\cite{su2017ferroelectric}, i.e., the state-of-the-art NVFF based checkpoint scheme, using
the mixture of Mediabench and MiBench
applications~\cite{lee1997mediabench,guthaus2001mibench,liu2017benchprime}. 
They were all compiled with standard -O3 optimization.
As a default configuration, \name uses the {\it SB} size of 40 entries with the
sequential search logic (Section~\ref{sec:alias})~\footnote{Since the target
microcontroller~\cite{MSP430FR5969} has 16 registers, the SB size must be at
least two times bigger than the register file size (16) to safely enable the
watchdog timer based checkpoint scheme shown in
Figure~\ref{fig:checkpoint}(b).}.
Table~\ref{table:comp} describes the hardware specifications of the baseline NVP and \name.
\ignore{
For the state-of-the-art work, we
referred to the prior works~\cite{gu2016nvpsim, su2017ferroelectric}; for \name,
we referred to the TI manual and the commodity hardware
design~\cite{WISP,MSP430FR5969}.  

For sensitivity analyses, we varied DMA
speed and NVM write/read latency ratio considering different
types of non-volatile memory technologies~\cite{elnawawy2017efficient,nair2015reducing,
hu2018persistence,xu2015overcoming}.

To set up DMA configuration is tricky because DMA must copy each data existing
in the proxy NVM buffer to the corresponding primary main memory locations, otherwise;
DMA may overwrite data to wrong memory location. To
achieve fine-grained data copy, \name uses the single DMA channel multiple
times in a row. Although a series of data copy seems to be not optimized,
DMA is still helpful because DMA engine can write the fine-grained data to NVM 
by about 4X faster than normal writes do~\cite{FRAMspeed}.
}

\begin{table}[h]
	\scriptsize
	\centering
	\begin{tabular}{|l|l|l|}
		\hline
		& \textbf{NVP}          & \textbf{\name}     \\ \hline \hline
		Capacitor        & 100nF        & 100nF/No  \\ \hline
		Computing Power     & 100uW/MHz    & 100uW/MHz \\ \hline
		Voltage Monitor(VM)  & 18uA    & No        \\ \hline
		Store Buffer & No           & Yes (Section~\ref{sec:alias})       \\ \hline
		DMA & No & Optional \\ \hline
		Von/Voff        & 3.3/2.8      & 1.8/1.8   \\ \hline
		Ckpt/Restore V   & 3.1/2.9      & No/1.8        \\ \hline
		Write/Read (latency)\footnotemark	&	120ns/20ns& 120ns/20ns	\\ \hline
		Write/Read (power) & 2mW & 2mW \\ \hline 
		Sleep/Wakeup T& 46/14us & 212/310us\cite{3us}        \\ \hline
		Recovery Point    & VM hit       & Boundary  \\ \hline
		ILP              & No           & Yes       \\ \hline
	\end{tabular}
	\caption{Simulation configuration}
	\label{table:comp}
	\vspace{-4mm}
\end{table}
\footnotetext{We configured the NVM write/read latency 
	based on the commodity design~\cite{MSP430FR5969} and 
the state-of-the-art works~\cite{xu2015overcoming,gu2016nvpsim}}

\begin{figure}[!htb]
\centering
\subfloat[Power trace\#1 (Home)]{\includegraphics[width=0.5\columnwidth, angle=0]{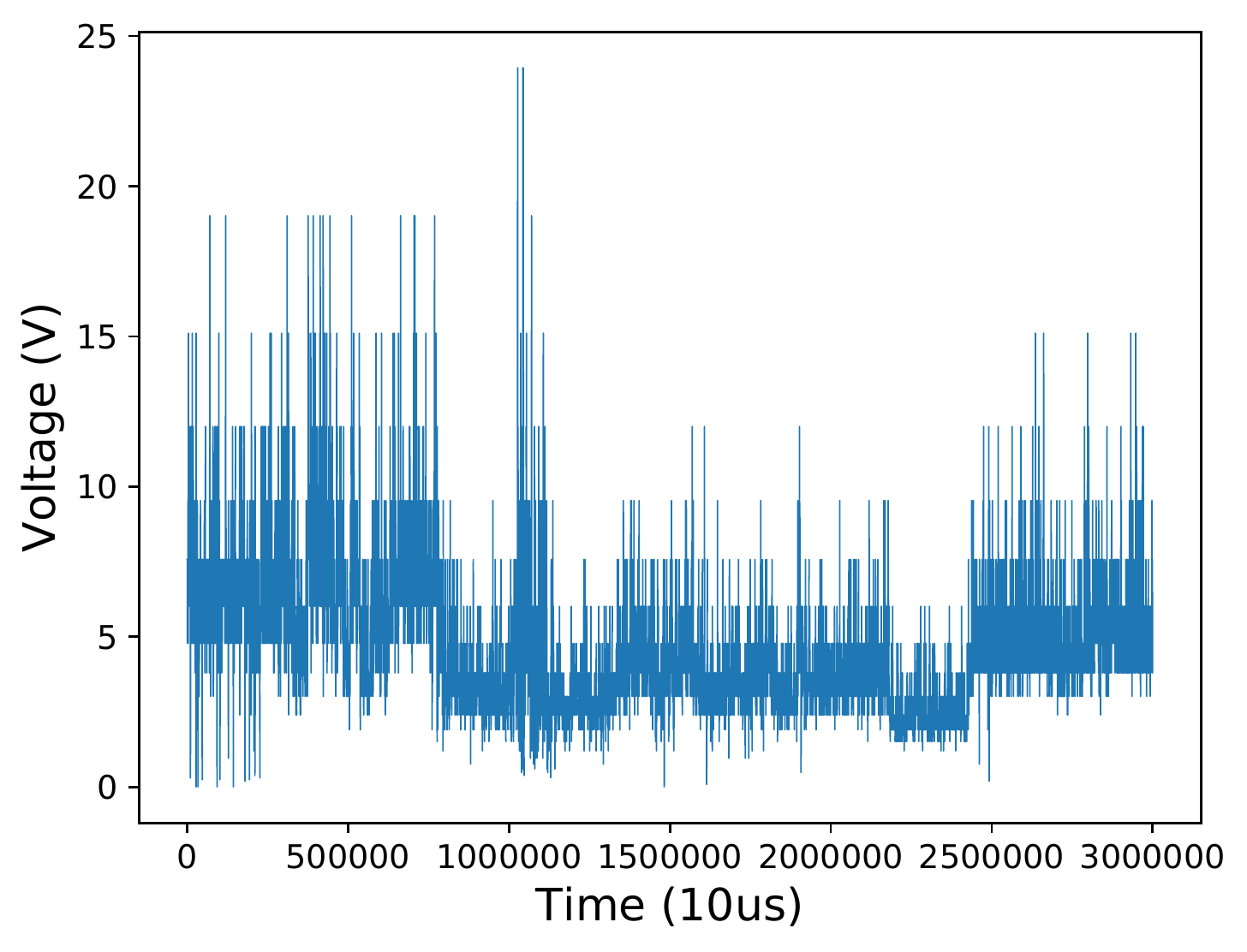}}
\subfloat[Power trace\#2 (Office)]{\includegraphics[width=0.5\columnwidth, angle=0]{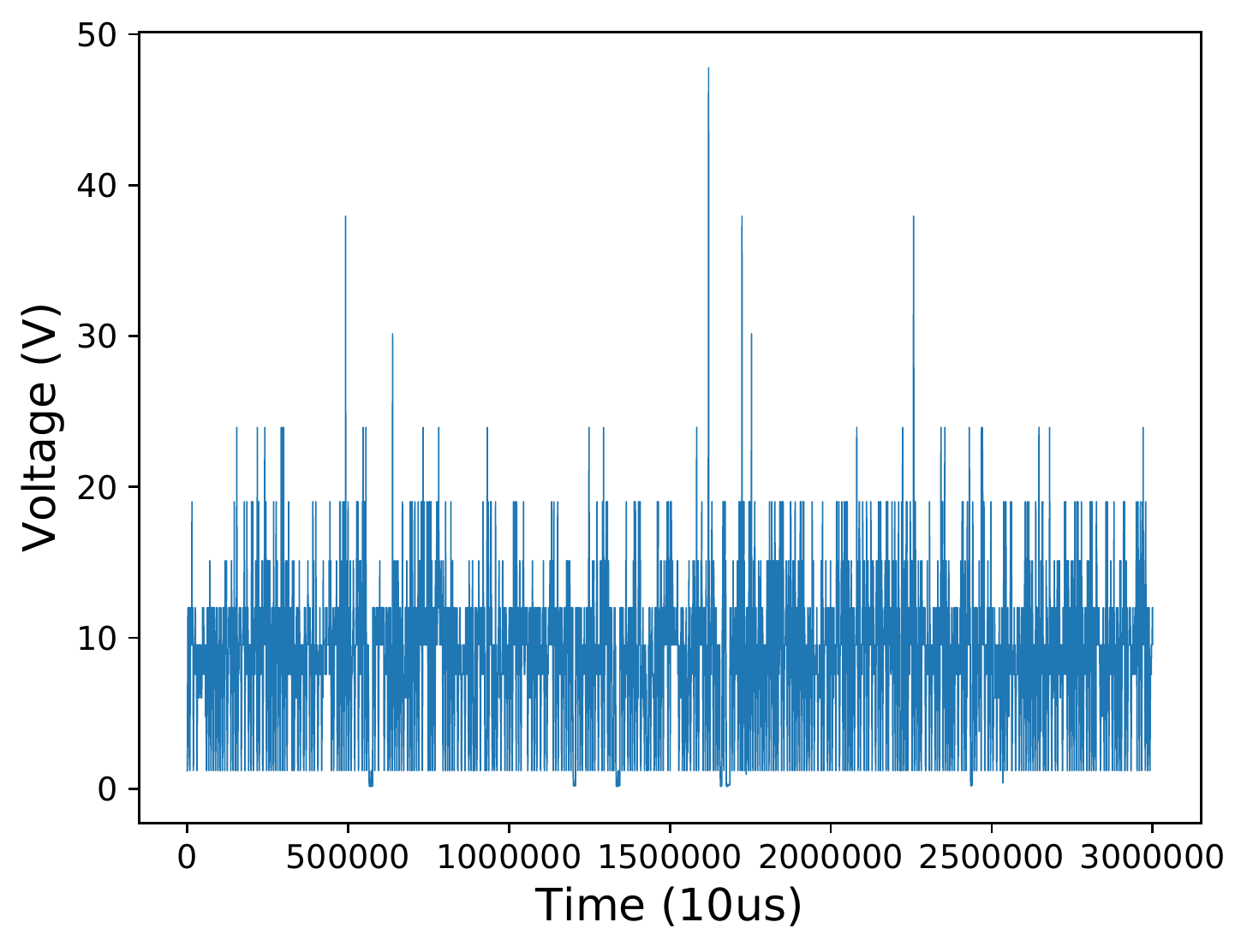}}
\caption{Energy harvesting voltage traces. Trace\#1 and\#2 incur $\approx$20
				and $\approx$400 power outages in every 30 seconds, respectively.}
\label{fig:vt}
\end{figure}

To evaluate \name for harsh environment with frequent power outages, 
we used two power traces of the NVP simulator which were
collected from real RF energy-harvesting systems~\cite{gu2016nvpsim}.
Figure~\ref{fig:vt} describes the shape of the two power traces: (a) home and
(b) office. In the following, we provide the detailed analyses of \name on
(1) hardware cost, (2) execution time with and without power failure,
and (3) energy consumption breakdown.

\subsection{Hardware Cost Analysis}
\begin{table}[h]
\footnotesize
\centering
\begin{tabular}{|c|cccc|}
	\hline 
Schemes & NVP~\cite{su2017ferroelectric} & Clank~\cite{Clank} & TCCP~\cite{TCCP} & \name \\ \hline
Core Type& InO & InO & OoO & InO \\
Buffers & No & 4 buffers & {\it SB} & {\it SB} \\
DMA & No & No & No & Optional \\ 
ISA Change & No & Yes & Yes & Optional \\ 
Double Backup & No & Yes & No & Yes \\ \hline
\textbf{Counter(s)} & No & No & \textbf{Yes (2)} & No \\
\textbf{NV Scratchpad} & No & \textbf{Yes} & No & No \\
\textbf{NVFF} & \textbf{Yes} & No & \textbf{Yes (+NVSB)} & No \\
\textbf{Extra Energy Buffer} & \textbf{Yes} & No & \textbf{Yes} & No \\
\textbf{Voltage Monitor} & \textbf{Yes} & No & \textbf{Yes} & No \\ \hline
Total Cost & High & High & High& Low \\ \hline
\end{tabular}
\caption{Hardware cost comparison: In the first column, the entries in bold are non-commodity hardware components, i.e., the bold marks represent expensive hardware modifications. Others have already been adopted to commodity hardware designs.}
\label{hardwarecost}
\vspace{-4mm}
\end{table}

This section analyzes the hardware cost of prior works~\cite{su2017ferroelectric, Clank, TCCP} and highlights the low
cost of \name. 
Table~\ref{hardwarecost} provides the major hardware cost comparison.
First,
NVP~\cite{su2017ferroelectric} requires the voltage monitor, NVFF, and extra energy buffer. 
The voltage monitor consumes a significant amount of energy and occupies
a nontrivial portion of die size~\cite{Clank}\footnote{
The die area occupied by the commodity voltage monitor is about
0.3$mm^2$~\cite{su2017ferroelectric}.}. Also, integrating the NVFF (nonvolatile
flip-flops), that must be laid out in close proximity to the volatile
flip-flops, in the core microarchitecture is complex and expensive due to the
manufacturing cost. Overall, the hardware cost of NVP is high. 
Second, Clank~\cite{Clank} introduces new hardware components such as
nonvolatile scratchpad and idempotence violation (i.e., antidependence) detector
with several memory buffers. Since the dependence tracking has to monitor every
single load/store and sweep the buffers for CAM based associative searches, it
is fair to say that the total cost of Clank is high.  Third, TCCP~\cite{TCCP}, a
variant of NVP, builds up an out-of-order processor. As with NVP, TCCP requires
the voltage monitor, extra energy buffer, and NVFF. In addition, TCCP
introduces a nonvolatile store buffer (NVSB) as well as two threshold counters
and their controller logic for varying the checkpoint interval. Given all this,
TCCP is another high cost approach. 

Finally, \name re-purposes the existing {\it SB} and introduces its 2-phase
release logic. Other than that, \name does not modify core microarchitecture
unlike above prior works. 
Although \name currently assumes a special load instruction for bypassing the
SB, this can be done without ISA change. The idea is to (1) set the least
significant bit of the aliased load address operand---which must be zero due to
the word granularity---and (2) let the pipeline architecture check the bit to
reset it and enable the SB bypassing.  Although the bit setting instruction
must be inserted at compile time, the overhead will not be significant thanks
to the small portion of aliased loads as shown in Figure~\ref{fig:aa}.
Although \name can opt for a DMA engine, it has already been adopted by
commodity in-order processors such as MSP430 series MCUs. Overall, the hardware
cost of \name is significantly lower than that of the prior works.


\begin{figure*}[!htb]
\centering
\centerline{\includegraphics[width=\textwidth, angle=0]{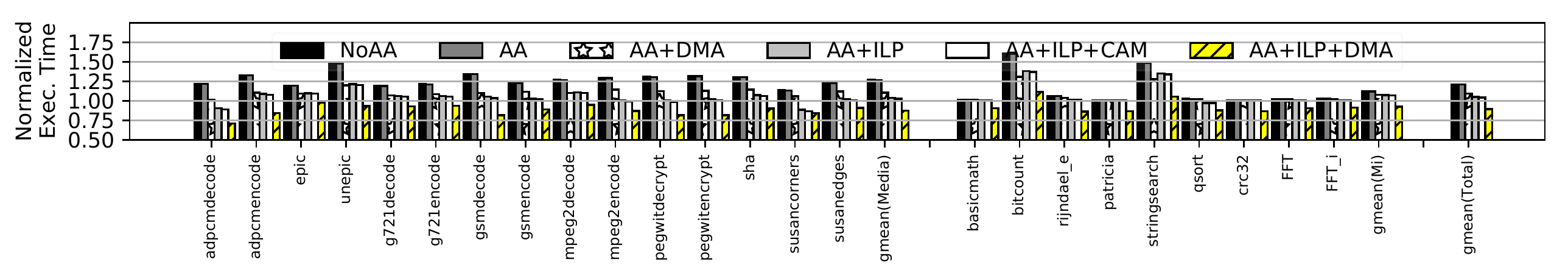}}
\centering
\caption{Normalized execution time of \name 
compared to NVP~\cite{su2017ferroelectric}.
As a default, \name enables SB bypass, ILP, and DMA support for all other experiments}.

\label{fig:performance}
\vspace{-4mm}
\end{figure*}
\begin{figure*}[!htb]
\centering
\centerline{\includegraphics[width=\textwidth, angle=0]{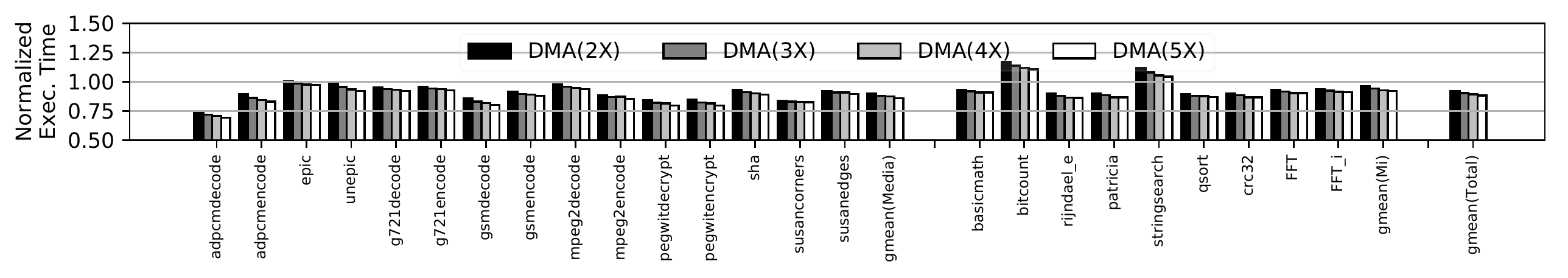}}
\centering
\caption{Normalized execution time of \name compared to
	NVP~\cite{su2017ferroelectric} varying DMA speed.
DMA(4X) is the default configuration for all other experiments.}
\label{fig:dmasense}
\vspace{-4mm}
\end{figure*}
\begin{figure*}[!htb]
\centering
\centerline{\includegraphics[width=\textwidth, angle=0]{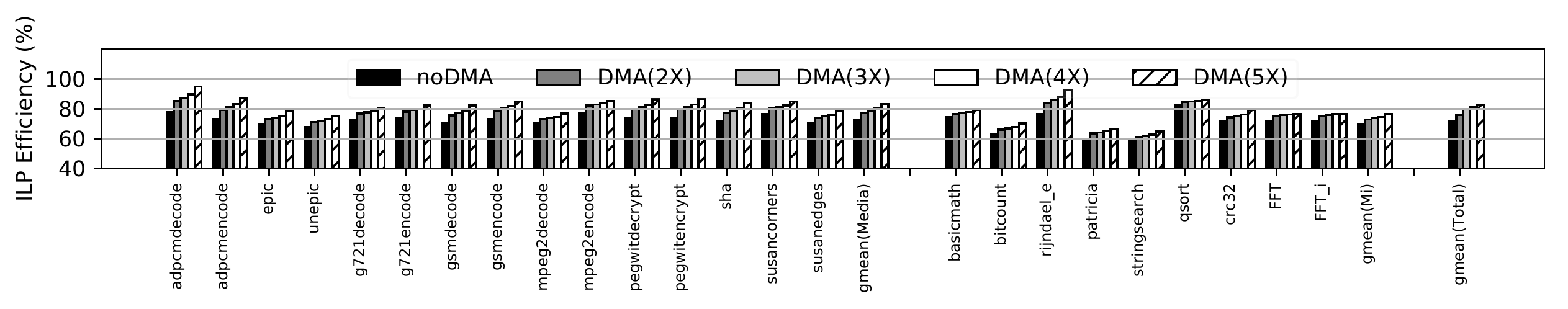}}
\centering
\caption{ILP Efficiency comparison varying DMA speed. DMA(4X) is the default configuration 
for all other experiments}
\label{fig:ILPeff}
\vspace{-4mm}
\end{figure*}

\ignore{
This graph has 6 different configurations: (1) no SB bypass(NoAA), (2) SB bypass(AA), 
(3) SB bypass and DMA support(AA+DMA), (4) SB bypass and ILP(AA+ILP), 
(5) SB bypass, ILP and CAM search(AA+ILP+CAM),
(6) SB bypass, ILP, and DMA support(AA+ILP+DMA).
}

\subsection{Execution Time Analysis with No Outage}
To analyze the execution time of \name, we first set the baseline to the
state-of-the-art NVP~\cite{su2017ferroelectric} with uninstrumented binaries.
We measured the execution time of \name for 24 benchmark applications 
with 6 configurations.
\ignore{ \todo{not sure what noAA means; is it no SB bypass? just say what the 6 cases
 in the figure mean, please} store buffer bypass (NoAA/AA), store buffer search
 mechanism (CAM/sequential), ILP configuration (NoILP/ILP), and DMA usage.
}

First, we analyzed the performance impact of the alias analysis based SB bypass by
turning it off (NoAA) and on (AA).  As shown in Figure~\ref{fig:performance},
without the {\it SB} bypass (NoAA), i.e., the first bar in the figure, \name
incurs about 24\% execution time overhead due to the region-based power failure
speculation overheads such as the 2-phase SB release and the inserted register
checkpoints.  When the SB bypass is enabled (AA), i.e., the second bar in
the figure, the resulting execution time reduction is only marginal.
This implies that the SB search is not the main source of the execution time
overhead.

Second, we also analyzed the impact of DMA and ILP on the execution time of applications.
Recall that DMA is used for fast memory-to-memory copy, 
and therefore it can only speed up the second phase of the SB release. 
When both SB bypass and DMA are enabled (AA+DMA), i.e., the third bar  
in Figure~\ref{fig:performance},
\name causes about 10\% execution time overhead; as with MSP430 microcontrollers, we set the DMA speed to 4X faster then normal memory copy as default.
When both SB bypass and ILP are enabled (AA+ILP), i.e., the fourth bar in the figure,
the resulting execution time overhead is only 4$\sim$5\% though DMA is not enabled. This confirms that ILP is the main reason for \name's high performance.

Finally, we enabled all the optimizations to see the performance bound of \name.
When the best configuration is set (AA+ILP+DMA), i.e., the sixth bar in the figure, 
\name rather outperforms the state-of-the-art NVP by 11\% on average. As the next section shows, the use of DMA is able to improve the ILP efficiency. In this way, \name can effectively hide the long latency of NVM writes involved in the 2-phase SB release. 

Interestingly, Figure~\ref{fig:performance} shows that
CAM search does not make a huge impact on the execution time on average.
When the CAM search is enabled with both SB bypass and ILP (AA+ILP+CAM),
i.e., the fifth bar in the figure,
there is only marginal difference compared to the sequential search with SB bypass and ILP (AA+ILP).
That is because 1$\sim$2\% of total loads access to the store buffer---as shown in Figure~\ref{fig:aa}---thanks to the precise alias analysis of \name's compiler. That is, only a few loads could get the CAM search benefit.
Note that all the other bars except for AA+ILP+CAM in Figure~\ref{fig:performance} use the sequential SB search logic.

\begin{figure*}[!htb]
\centering
\centerline{\includegraphics[width=\textwidth, angle=0]{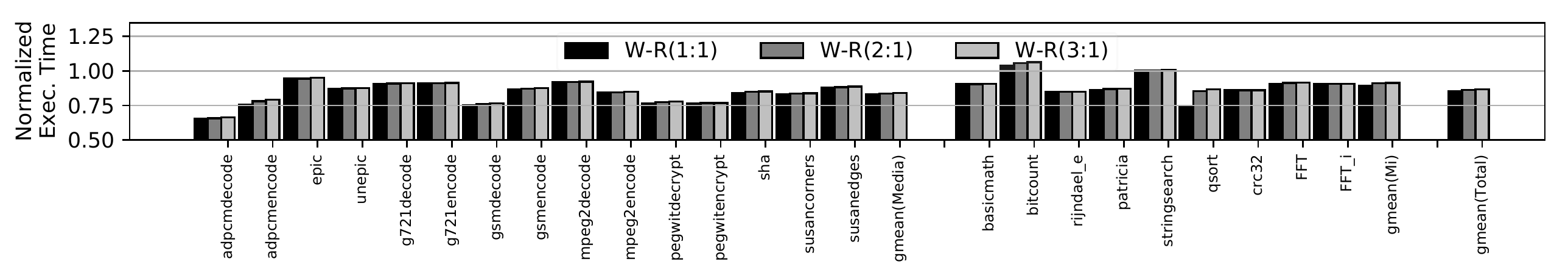}}
\centering
\caption{Normalized execution time of \name compared to NVP
~\cite{su2017ferroelectric} varying the write-to-read ratio of NVM. The ratio, 6:1, is the default configuration for all other experiments.}
\label{fig:writeread}
\vspace{-4mm}
\end{figure*}
\begin{figure*}[!htb]
\centering
				\subfloat[Power trace\#1 (Home)]{\includegraphics[width=\columnwidth, angle=0]{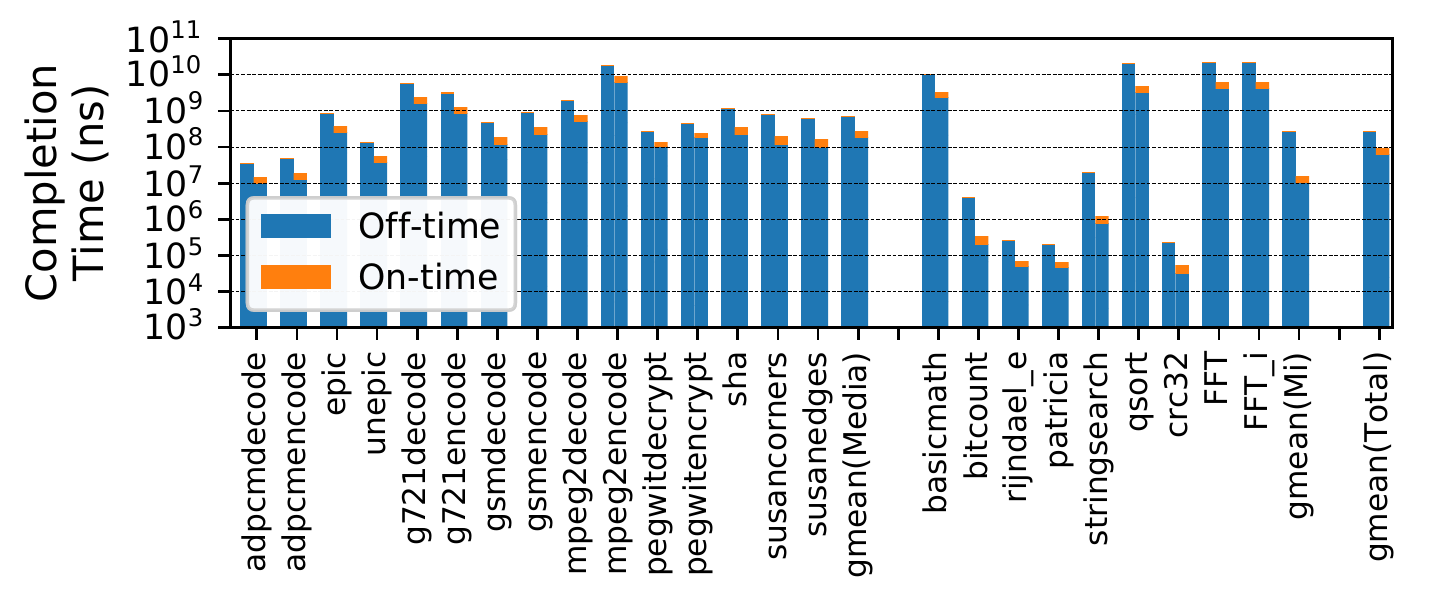}}
				\subfloat[Power trace\#2 (Office)]{\includegraphics[width=\columnwidth, angle=0]{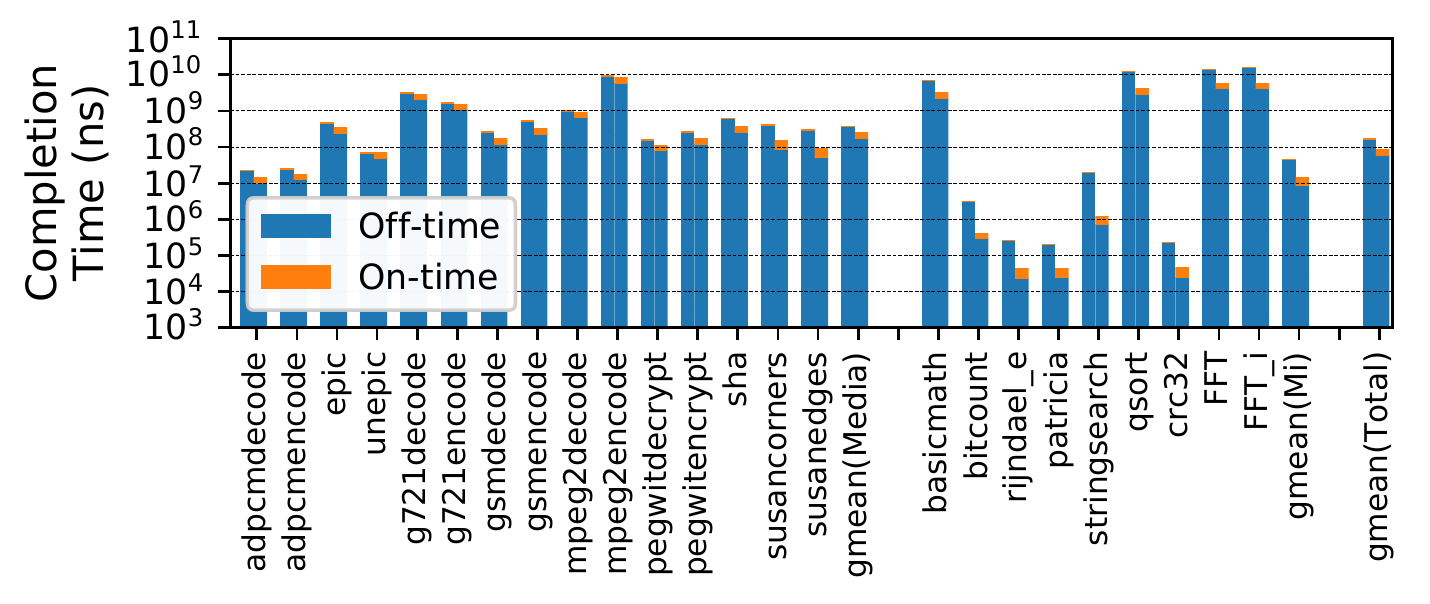}}
\caption{Completion time comparison. The 1st/2nd bars of each application
				represent the times of NVP and \name, respectively.}
\label{fig:intermittent}
\vspace{-4mm}
\end{figure*}

\subsubsection{Sensitivity Analysis}
\label{sec:sensitivity}
We explored the performance impact of the DMA and NVM technology with
the highly optimized \name (AA+ILP+DMA).
First, we varied the DMA speed of data transfer in NVM, i.e., 2X, 3X, and 5X faster than a normal NVM copy---and then measured the resulting execution times of NVP and \name for the same set of benchmark applications.

Figure~\ref{fig:dmasense} shows the normalized execution time of \name compared to NVP which is the same baseline used in the prior experiment.
To a large extent, \name becomes faster as the DMA speed is increased.
When the DMA speed is 5X, \name can achieve $\sim$6\% speedup than the default speed of 4X.

To further analyze the correlation between the DMA speed and ILP execution, 
we measured the ILP efficiency varying the DMA speed.
\ignore{
by using a cost model below:
$Eff = min(T(region),T(backup))/T(backup)$~\footnote{
	$T()$ represents the execution time of a given entity.
}.
}	
The ILP efficiency is defined as how much the time taken for the 2-phase SB release of a code region is overlapped with the execution time of the next region. 
For example, if the SB release time is completely overlapped with the next region execution, 
the ILP efficiency is 100\%. Note that the perfect efficiency is achieved when the region execution time is greater than or equal to the SB release time; either way, the SB release time is fully hidden, the ILP efficiency is 100\%.
On the other hand, if the next region finishes while the SB release is still pending, the ILP efficiency is decreased. That is because \name must wait---at the end of the next region---for the SB release to finish.
As shown in Figure~\ref{fig:ILPeff}, 70$\sim$82\% of the 2-phase SB release can be overlapped with the next region execution when the DMA speed is 2X$\sim$5X. 
\ignore{
The ILP efficiency somehow slightly improves when the DMA speed increases.
Since more than 40\% of program code regions are
already long enough to be overlapped, they may not
affect to ILP efficiency no matter how fast DMA speed is.
}

\begin{table}[h]
\small
\centering
\begin{tabular}{|l|l|l|l|l|}
\hline
 Memory & FRAM~\cite{MSP430FR5994} & NVsim~\cite{elnawawy2017efficient} & PCM~\cite{nair2015reducing,hu2018persistence} & Re-RAM~\cite{xu2015overcoming} \\ \hline
Ratio & 1:1 & 2:1 & 3:1 & 6:1 \\ \hline
\end{tabular}
\caption{Write-to-read ratios of different NVM technologies.}
\label{table:NVM}
\vspace{-4mm}
\end{table}

Second, we varied the NVM write/read latency ratio, i.e., 1:1, 2:1, and
3:1, assuming different NVM technologies\cite{elnawawy2017efficient,nair2015reducing,
hu2018persistence,xu2015overcoming} shown in Table~\ref{table:NVM}.
Figure~\ref{fig:writeread} shows
the normalized execution times of \name compared to the same baseline NVP again.
On average, \name outperforms the NVP by about 13$\sim$16\% 
when the NVM write/read ratio becomes 3$\sim$1:1.



\ignore{
\begin{figure}[!h]
\centering
\centerline{\includegraphics[width=\columnwidth, angle=0]{fig/accum.pdf}}
\centering
\caption{Region characteristics for ILP efficiency}
\label{fig:ILPregion}
\end{figure}
}

\ignore{
Figure~\ref{fig:ILPregion} describes the program region characteristics
corresponding to ILP efficiently.
Interestingly, qsort application mostly has long regions, i.e.,
80\% of regions are already long enough to overlap the 2-phase backup latency.
Due to the long regions, ILP+DMA efficiency improved only 3$\sim$4\% compared to
ILP only.
On the other hand, adpcmdecode and rijndael applications has optimal size regions
for ILP benefit, e.g., 95\% of total regions 
}



\subsection{Execution Time Analysis with Outages}

To test the ability to make forward execution progress in the presence
of a myriad of power outages,
we measured the completion time of benchmark applications using
two voltage traces shown in Figure~\ref{fig:vt}; they are 
collected from a real RF-based energy harvesting system when it is deployed in home (a) and office (a).
Figure~\ref{fig:intermittent} shows the completion time of the baseline NVP (the first bar) and \name (the second bar) with breaking down the time to 2 parts, i.e., power-off-time and power-on-time.
As shown in the figure, the system off-time dominates the completion time
of both NVP and \name.  However, NVP is designed to wake up
at 1.5$\sim$3X higher voltage level than the minimum supply voltage of MCUs,
due to voltage monitor issues (see Section~\ref{sec:hardwareprior}).
This implies that NVP should stay in a sleep mode for a substantial amount of time without making forward progress.
Unlike the NVP, \name can start to operate once the minimum supply voltage is secured,
thus achieving further forward progress. 
Figure~\ref{fig:intermittent} (a) and (b) highlights that 
\name outperforms the NVP by 3.0X and 1.8X in the trace\#1 (a) and trace\#2 (a), respectively.

Interestingly, the NVP makes further forward execution progress in trace\#1 than trace\#2.
As shown in Figure~\ref{fig:intermittent} (b), NVP's completion time using trace\#1 is only 60\% of that of using trace\#1. Given that trace\#2 has relatively less power outages than trace\#1, NVP tends to prefer more reliable voltage trace.
With that in mind, we expect that \name can outperform the NVP more significantly when the energy source is more unreliable.

\subsection{Energy Breakdown with Outages}
\label{sec:break}
Finally, we analyzed the average energy consumption breakdown across power
outages~\footnote{\name shows similar energy consumption trends in both power
traces. On average, the total energy consumption of \name in the presence
of power failures using two traces is 2$\sim$3X less than the NVP's.}.
The total energy consumption can be divided into two parts: one under ILP execution and the other under non-ILP execution.
The ILP part is further broken down to successful- and mis-speculation, each of which is comprised of 3 parts: the Phase1/Phase2 of the SB release and the computation. On the other hand, the non-ILP part is two-fold: NoILP and re-execution. NoILP is simply the energy consumption of \name when it executes without ILP excluding that of re-executing any interrupted regions. 

\begin{figure}[!h]
\centering
\centerline{\includegraphics[width=\columnwidth, angle=0]{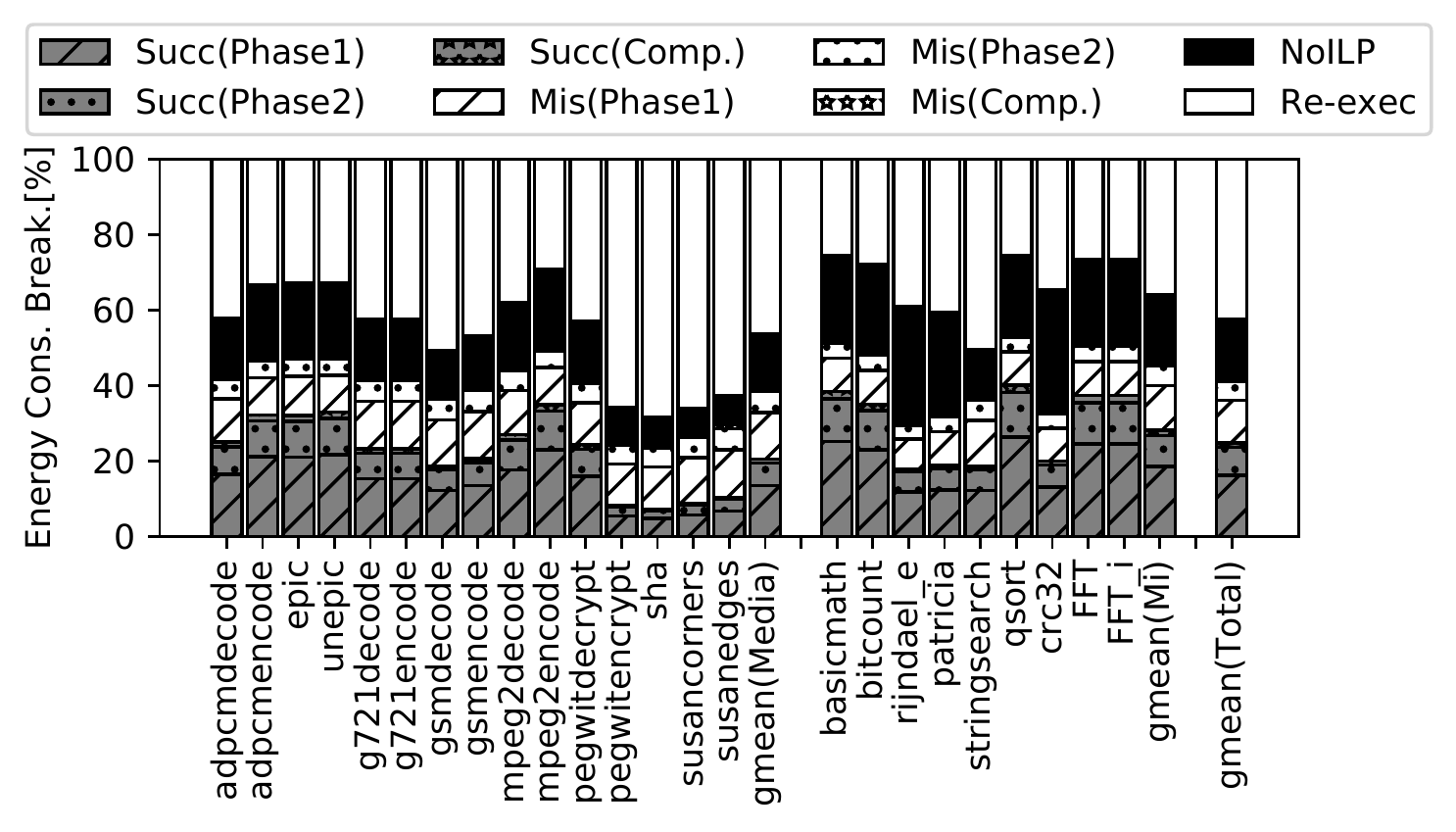}}
\centering
\caption{Energy consumption breakdown of \name
}

\label{fig:energybreak}
\end{figure}

Figure~\ref{fig:energybreak} indicates that the overhead of \name mostly comes
from the re-execution cost---i.e., Re-exec in the figure.  Although \name
enables the ILP and the watchdog-timer-based checkpoint in an adaptive manner
according to a dynamic power failure pattern, the adaptation may not help for
the first time to make progress (see Section~\ref{sec:forward}).  For example,
to avoid stagnation, \name might need to perform multiple times of the
adaptation in a reactive manner (involving the sequence of ILP off -> watchdog
timer on -> timer halving). Thus, the re-execution consumes the harvested
energy without making actual progress until \name finally gets out of the
stagnating region after the multiple adaptations.  As shown in
Figure~\ref{fig:energybreak}, the re-execution consumes 40\% of the total
energy on average.

On average, \name consumes about 40\% of its total harvested energy for ILP
executions while it does the rest of the energy for non-ILP executions.  NoILP
consumes 20\% of total energy on average due to the adaptation of \name which
throttles down its execution to escape (potentially) stagnating regions.
In particular, extra NVM writes (i.e., the Phase2 of the SB release during successful- and
mis-speculation) account for 18\% of the total energy consumption on average. Also, it turns out that the wasted energy of mis-speculated execution (i.e., computation during
mis-speculation) is negligible thanks to \name's adaptive execution.



\section{Other Related Work} \label{sec:Rel} 

The problem of ensuring data consistency and improving the forward 
progress of an intermittently powered system is at the heart of energy
harvesting computing. A variety of different techniques have been proposed, but they end
up introducing non-traditional microarchitecture modifications or incurring significant performance overheads.
In contrast, \name can achieve high performance speculative intermittent computation
for commodity energy-harvesting microcontrollers without expensive modifications.

Ma \emph {et al} propose so-called incidental computing~\cite{ma2017incidental}.
This scheme attempts to trade off the output quality of a program to improve forward execution progress. The key observation is that input data for many
signal/image processing applications are mostly free from data dependencies, and
the applications often contain memory independent loop iterations that could be
skipped over in their entirety.  With that in mind, the scheme ignores some
computations interrupted due to power failure. When power comes back, it starts
to process the most recent data, producing an earlier output at the expense of
the quality degradation.  Unfortunately, the scheme requires user-intervention,
e.g., programmers must mark the skip points, recovery points, and so on.  Unlike
the scheme, \name does not require user intervention at all but can still
achieve the same goal, i.e., improving forward progress, without compromising
the output quality. 

Ma \emph {et al} also suggest a hybrid processor equipped with both in-order and
out-of-order pipelines to adapt the processor execution to underlying energy
harvesting condition~\cite{ma2017spendthrift}.  This scheme chooses one of the
pipelines according to run-time power failure behaviors.  Unfortunately, the
scheme does not only require expensive hardware modifications due to the hybrid
processor design but also cause additional switching delay and energy
consumption when it turns on/off the hardware resources.  In contrast, \name
builds on top of the current commodity in-order microarchitecture without
expensive hardware modifications. Nevertheless, \name can achieve further
forward execution progress with the help of speculative intermittent
computation.

Colins \emph {et al} propose a reconfigurable energy buffer with multiple
capacitor banks for energy efficiency~\cite{colin2018reconfigurable}.  Since
each application may require a different amount of energy, 
the authors attempt to reconfigure the energy capacity to match the
application's demand by switching on/off some part of the banks.  However, the
energy buffer can suffer from unstable discharge due to in-field aging problems
or environmental changes. Thus, the scheme may be unreliable especially when
energy harvesting systems are deployed in harsh environment. While the
capacitor reconfiguration techniques are
orthogonal to those of \name, it can achieve a truly reliable crash consistency  without the custom capacitor logic and the additional hardware support.

Ruppel \emph {et al} devise an event-driven transactional concurrency control 
scheme. To provide a transaction, that can include a set of tasks, with
failure-atomicity, the authors leverage interrupt service
routines~\cite{ruppel2019transactional}. For this purpose, program tasks
must be delineated in the first place and encapsulated in a transaction and an
asynchronous event. Apart from this burden placed on programmers, the scheme
causes a significant performance overhead.
In contrast, \name is an automated compiler/architecture co-design scheme that enables high-performance intermittent computation.


\section{Summary} This paper presents \name, an architecture/compiler co-designed scheme, that can
work for commodity in-order processors, to achieve low-cost yet performant
intermittent computation.  \name takes advantage of power failure speculation to
enable crash consistency without significant hardware and performance overheads.
In particular, \name realizes instruction level parallelism on top of the
in-order processor pipeline to hide the long latency of nonvolatile memory writes, thereby
improving the performance significantly.  Our experiments on a real energy
harvesting trace with frequent power outages demonstrate that \name outperforms
the state-of-the-art nonvolatile processor across a variety of benchmark
applications by 3X on average.

\begin{acks}
The original article was presented in MICRO’19~\cite{choi2019cospec}.
We appreciate anonymous reviewers for their constructive comments.
This work was supported by NSF grants 1750503 (CAREER Award) and 1814430.
\end{acks}

\bibliographystyle{ACM-Reference-Format}
\balance
\bibliography{refs,refs2,references}


\begin{thebibliography}{79}


\ifx \showCODEN    \undefined \def \showCODEN     #1{\unskip}     \fi
\ifx \showDOI      \undefined \def \showDOI       #1{#1}\fi
\ifx \showISBNx    \undefined \def \showISBNx     #1{\unskip}     \fi
\ifx \showISBNxiii \undefined \def \showISBNxiii  #1{\unskip}     \fi
\ifx \showISSN     \undefined \def \showISSN      #1{\unskip}     \fi
\ifx \showLCCN     \undefined \def \showLCCN      #1{\unskip}     \fi
\ifx \shownote     \undefined \def \shownote      #1{#1}          \fi
\ifx \showarticletitle \undefined \def \showarticletitle #1{#1}   \fi
\ifx \showURL      \undefined \def \showURL       {\relax}        \fi
\providecommand\bibfield[2]{#2}
\providecommand\bibinfo[2]{#2}
\providecommand\natexlab[1]{#1}
\providecommand\showeprint[2][]{arXiv:#2}

\bibitem[\protect\citeauthoryear{??}{FRA}{2015}]%
        {FRAMspeed}
 \bibinfo{year}{2015}\natexlab{}.
\newblock \bibinfo{title}{Maximizing Write Speed on the MSP430 FRAM}.
\newblock
  \bibinfo{howpublished}{\url{http://www.ti.com/mcu/docs/litabsmultiplefilelist.tsp?sectionId=96&tabId=1502&literatureNumber=slaa498b&docCategoryId=1&familyId=5012}}.
\newblock
\newblock
\shownote{Accessed: 2018-10-14.}


\bibitem[\protect\citeauthoryear{??}{MSP}{2016}]%
        {MSP430FR5994}
 \bibinfo{year}{2016}\natexlab{}.
\newblock \bibinfo{title}{MSP430FR5994LaunchPad Development Kit
  (MSPEXP430FR5994)}.
\newblock
  \bibinfo{howpublished}{\url{http://www.ti.com/lit/ug/slau678a/slau678a.pdf}}.
\newblock
\newblock
\shownote{Accessed: 2017-11-08.}


\bibitem[\protect\citeauthoryear{??}{min}{2017}]%
        {minorCPU}
 \bibinfo{year}{2017}\natexlab{}.
\newblock \bibinfo{title}{ARM Research Starter Kit: System Modeling Using
  gem5}.
\newblock
  \bibinfo{howpublished}{\url{https://raw.githubusercontent.com/arm-university/arm-gem5-rsk/master/gem5_rsk.pdf}}.
\newblock
\newblock
\shownote{Accessed: 2018-11-18.}


\bibitem[\protect\citeauthoryear{??}{MSP}{2017}]%
        {MSP430FR5969}
 \bibinfo{year}{2017}\natexlab{}.
\newblock \bibinfo{title}{MSP430FR59xx Mixed-Signal Microcontrollers (Rev. F)}.
\newblock
  \bibinfo{howpublished}{\url{http://www.ti.com/lit/ds/symlink/msp430fr5969.pdf}}.
\newblock
\newblock
\shownote{Accessed: 2017-11-08.}


\bibitem[\protect\citeauthoryear{Aantjes, Majid, and Pawe{\l}czak}{Aantjes
  et~al\mbox{.}}{2016}]%
        {aantjes2016testbed}
\bibfield{author}{\bibinfo{person}{Henko Aantjes}, \bibinfo{person}{Amjad~Y
  Majid}, {and} \bibinfo{person}{Przemys{\l}aw Pawe{\l}czak}.}
  \bibinfo{year}{2016}\natexlab{}.
\newblock \showarticletitle{A Testbed for Transiently Powered Computers}.
\newblock \bibinfo{journal}{\emph{arXiv preprint arXiv:1606.07623}}
  (\bibinfo{year}{2016}).
\newblock


\bibitem[\protect\citeauthoryear{Andersen}{Andersen}{1994}]%
        {andersen1994program}
\bibfield{author}{\bibinfo{person}{Lars~Ole Andersen}.}
  \bibinfo{year}{1994}\natexlab{}.
\newblock \emph{\bibinfo{title}{Program analysis and specialization for the C
  programming language}}.
\newblock \bibinfo{thesistype}{Ph.D. Dissertation}. \bibinfo{school}{University
  of Cophenhagen}.
\newblock


\bibitem[\protect\citeauthoryear{Baghsorkhi and Margiolas}{Baghsorkhi and
  Margiolas}{2018}]%
        {baghsorkhi2018automating}
\bibfield{author}{\bibinfo{person}{Sara~S Baghsorkhi} {and}
  \bibinfo{person}{Christos Margiolas}.} \bibinfo{year}{2018}\natexlab{}.
\newblock \showarticletitle{Automating efficient variable-grained resiliency
  for low-power IoT systems}. In \bibinfo{booktitle}{\emph{Proceedings of the
  2018 International Symposium on Code Generation and Optimization}}. ACM,
  \bibinfo{pages}{38--49}.
\newblock


\bibitem[\protect\citeauthoryear{Balsamo, Weddell, Das, Arreola, Brunelli,
  Al-Hashimi, Merrett, and Benini}{Balsamo et~al\mbox{.}}{2016}]%
        {balsamo2016hibernus++}
\bibfield{author}{\bibinfo{person}{Domenico Balsamo}, \bibinfo{person}{Alex~S
  Weddell}, \bibinfo{person}{Anup Das}, \bibinfo{person}{Alberto~Rodriguez
  Arreola}, \bibinfo{person}{Davide Brunelli}, \bibinfo{person}{Bashir~M
  Al-Hashimi}, \bibinfo{person}{Geoff~V Merrett}, {and} \bibinfo{person}{Luca
  Benini}.} \bibinfo{year}{2016}\natexlab{}.
\newblock \showarticletitle{Hibernus++: a self-calibrating and adaptive system
  for transiently-powered embedded devices}.
\newblock \bibinfo{journal}{\emph{IEEE Transactions on Computer-Aided Design of
  Integrated Circuits and Systems}} \bibinfo{volume}{35}, \bibinfo{number}{12}
  (\bibinfo{year}{2016}), \bibinfo{pages}{1968--1980}.
\newblock


\bibitem[\protect\citeauthoryear{Beeby and White}{Beeby and White}{2014}]%
        {beeby2014}
\bibfield{author}{\bibinfo{person}{S. Beeby} {and} \bibinfo{person}{N. White}.}
  \bibinfo{year}{2014}\natexlab{}.
\newblock \bibinfo{booktitle}{\emph{Energy Harvesting for Autonomous Systems}}.
\newblock \bibinfo{publisher}{Artech House, Incorporated}.
\newblock
\showISBNx{9781596937192}
\urldef\tempurl%
\url{https://books.google.fr/books?id=7H9xdFd4sikC}
\showURL{%
\tempurl}


\bibitem[\protect\citeauthoryear{Binkert, Beckmann, Black, Reinhardt, Saidi,
  Basu, Hestness, Hower, Krishna, Sardashti, Sen, Sewell, Shoaib, Vaish, Hill,
  and Wood}{Binkert et~al\mbox{.}}{2011}]%
        {Binkert11gem5}
\bibfield{author}{\bibinfo{person}{Nathan Binkert}, \bibinfo{person}{Bradford
  Beckmann}, \bibinfo{person}{Gabriel Black}, \bibinfo{person}{Steven~K.
  Reinhardt}, \bibinfo{person}{Ali Saidi}, \bibinfo{person}{Arkaprava Basu},
  \bibinfo{person}{Joel Hestness}, \bibinfo{person}{Derek~R. Hower},
  \bibinfo{person}{Tushar Krishna}, \bibinfo{person}{Somayeh Sardashti},
  \bibinfo{person}{Rathijit Sen}, \bibinfo{person}{Korey Sewell},
  \bibinfo{person}{Muhammad Shoaib}, \bibinfo{person}{Nilay Vaish},
  \bibinfo{person}{Mark~D. Hill}, {and} \bibinfo{person}{David~A. Wood}.}
  \bibinfo{year}{2011}\natexlab{}.
\newblock \showarticletitle{The Gem5 Simulator}.
\newblock \bibinfo{journal}{\emph{SIGARCH Comput. Archit. News}}
  \bibinfo{volume}{39}, \bibinfo{number}{2} (\bibinfo{date}{Aug.}
  \bibinfo{year}{2011}).
\newblock


\bibitem[\protect\citeauthoryear{Campbell, Ghena, and Dutta}{Campbell
  et~al\mbox{.}}{2014}]%
        {CampbellGD14}
\bibfield{author}{\bibinfo{person}{Bradford Campbell}, \bibinfo{person}{Branden
  Ghena}, {and} \bibinfo{person}{Prabal Dutta}.}
  \bibinfo{year}{2014}\natexlab{}.
\newblock \showarticletitle{Energy-harvesting thermoelectric sensing for
  unobtrusive water and appliance metering}. In
  \bibinfo{booktitle}{\emph{Proceedings of the 2nd International Workshop on
  Energy Neutral Sensing Systems, ENSsys '14, Memphis, Tennessee, USA, November
  6, 2014}}. \bibinfo{pages}{7--12}.
\newblock
\urldef\tempurl%
\url{https://doi.org/10.1145/2675683.2675692}
\showDOI{\tempurl}


\bibitem[\protect\citeauthoryear{Choi, Joe, Kim, and Jung}{Choi
  et~al\mbox{.}}{2019a}]%
        {choi2019achieving}
\bibfield{author}{\bibinfo{person}{Jongouk Choi}, \bibinfo{person}{Hyunwoo
  Joe}, \bibinfo{person}{Yongjoo Kim}, {and} \bibinfo{person}{Changhee Jung}.}
  \bibinfo{year}{2019}\natexlab{a}.
\newblock \showarticletitle{Achieving Stagnation-Free Intermittent Computation
  with Boundary-Free Adaptive Execution}. In \bibinfo{booktitle}{\emph{2019
  IEEE Real-Time and Embedded Technology and Applications Symposium (RTAS)}}.
  IEEE, \bibinfo{pages}{331--344}.
\newblock


\bibitem[\protect\citeauthoryear{Choi, Liu, and Jung}{Choi
  et~al\mbox{.}}{2019b}]%
        {choi2019cospec}
\bibfield{author}{\bibinfo{person}{Jongouk Choi}, \bibinfo{person}{Qingrui
  Liu}, {and} \bibinfo{person}{Changhee Jung}.}
  \bibinfo{year}{2019}\natexlab{b}.
\newblock \showarticletitle{CoSpec: Compiler Directed Speculative Intermittent
  Computation}. In \bibinfo{booktitle}{\emph{Proceedings of the 52nd Annual
  IEEE/ACM International Symposium on Microarchitecture}}. ACM,
  \bibinfo{pages}{399--412}.
\newblock


\bibitem[\protect\citeauthoryear{Colin and Lucia}{Colin and Lucia}{2015}]%
        {Chain}
\bibfield{author}{\bibinfo{person}{Alexei Colin} {and} \bibinfo{person}{Brandon
  Lucia}.} \bibinfo{year}{2015}\natexlab{}.
\newblock \showarticletitle{Chain: Tasks and Channels for Reliable Intermittent
  Programs.}. In \bibinfo{booktitle}{\emph{In Proceedings of the 2016 ACM
  SIGPLAN International Conference on Object-Oriented Programming, Systems,
  Languages, and Applications (OOPSLA)}}. ACM, \bibinfo{pages}{514--530}.
\newblock


\bibitem[\protect\citeauthoryear{Colin, Ruppel, and Lucia}{Colin
  et~al\mbox{.}}{2018}]%
        {colin2018reconfigurable}
\bibfield{author}{\bibinfo{person}{Alexei Colin}, \bibinfo{person}{Emily
  Ruppel}, {and} \bibinfo{person}{Brandon Lucia}.}
  \bibinfo{year}{2018}\natexlab{}.
\newblock \showarticletitle{A Reconfigurable Energy Storage Architecture for
  Energy-harvesting Devices}. In \bibinfo{booktitle}{\emph{Proceedings of the
  Twenty-Third International Conference on Architectural Support for
  Programming Languages and Operating Systems}}. ACM,
  \bibinfo{pages}{767--781}.
\newblock


\bibitem[\protect\citeauthoryear{de~Kruijf and Sankaralingam}{de~Kruijf and
  Sankaralingam}{2013}]%
        {de2013idempotent}
\bibfield{author}{\bibinfo{person}{Marc de Kruijf} {and}
  \bibinfo{person}{Karthikeyan Sankaralingam}.}
  \bibinfo{year}{2013}\natexlab{}.
\newblock \showarticletitle{Idempotent code generation: Implementation,
  analysis, and evaluation}. In \bibinfo{booktitle}{\emph{Code Generation and
  Optimization (CGO), 2013 IEEE/ACM International Symposium on}}. IEEE,
  \bibinfo{pages}{1--12}.
\newblock


\bibitem[\protect\citeauthoryear{de~Kruijf, Sankaralingam, and Jha}{de~Kruijf
  et~al\mbox{.}}{2012}]%
        {de2012static}
\bibfield{author}{\bibinfo{person}{Marc~A. de Kruijf},
  \bibinfo{person}{Karthikeyan Sankaralingam}, {and} \bibinfo{person}{Somesh
  Jha}.} \bibinfo{year}{2012}\natexlab{}.
\newblock \showarticletitle{Static Analysis and Compiler Design for Idempotent
  Processing}. In \bibinfo{booktitle}{\emph{Proceedings of the 33rd ACM SIGPLAN
  Conference on Programming Language Design and Implementation}}
  \emph{(\bibinfo{series}{PLDI '12})}. \bibinfo{publisher}{ACM},
  \bibinfo{address}{New York, NY, USA}, \bibinfo{pages}{475--486}.
\newblock
\showISBNx{978-1-4503-1205-9}
\urldef\tempurl%
\url{https://doi.org/10.1145/2254064.2254120}
\showDOI{\tempurl}


\bibitem[\protect\citeauthoryear{Elnawawy, Alshboul, Tuck, and
  Solihin}{Elnawawy et~al\mbox{.}}{2017}]%
        {elnawawy2017efficient}
\bibfield{author}{\bibinfo{person}{Hussein Elnawawy}, \bibinfo{person}{Mohammad
  Alshboul}, \bibinfo{person}{James Tuck}, {and} \bibinfo{person}{Yan
  Solihin}.} \bibinfo{year}{2017}\natexlab{}.
\newblock \showarticletitle{Efficient Checkpointing of Loop-Based Codes for
  Non-Volatile Main Memory}. In \bibinfo{booktitle}{\emph{2017 26th
  International Conference on Parallel Architectures and Compilation Techniques
  (PACT)}}. IEEE, \bibinfo{pages}{318--329}.
\newblock


\bibitem[\protect\citeauthoryear{Gu, Liu, Wang, Li, and Yang}{Gu
  et~al\mbox{.}}{2016}]%
        {gu2016nvpsim}
\bibfield{author}{\bibinfo{person}{Yizi Gu}, \bibinfo{person}{Yongpan Liu},
  \bibinfo{person}{Yiqun Wang}, \bibinfo{person}{Hehe Li}, {and}
  \bibinfo{person}{Huazhong Yang}.} \bibinfo{year}{2016}\natexlab{}.
\newblock \showarticletitle{NVPsim: A simulator for architecture explorations
  of nonvolatile processors}. In \bibinfo{booktitle}{\emph{Design Automation
  Conference (ASP-DAC), 2016 21st Asia and South Pacific}}. IEEE,
  \bibinfo{pages}{147--152}.
\newblock


\bibitem[\protect\citeauthoryear{Gunadi and Lipasti}{Gunadi and
  Lipasti}{2007}]%
        {gunadi2007position}
\bibfield{author}{\bibinfo{person}{Erika Gunadi} {and} \bibinfo{person}{Mikko~H
  Lipasti}.} \bibinfo{year}{2007}\natexlab{}.
\newblock \showarticletitle{A position-insensitive finished store buffer}. In
  \bibinfo{booktitle}{\emph{Computer Design, 2007. ICCD 2007. 25th
  International Conference on}}. IEEE, \bibinfo{pages}{105--112}.
\newblock


\bibitem[\protect\citeauthoryear{Guthaus, Ringenberg, Ernst, Austin, Mudge, and
  Brown}{Guthaus et~al\mbox{.}}{2001}]%
        {guthaus2001mibench}
\bibfield{author}{\bibinfo{person}{Matthew~R Guthaus},
  \bibinfo{person}{Jeffrey~S Ringenberg}, \bibinfo{person}{Dan Ernst},
  \bibinfo{person}{Todd~M Austin}, \bibinfo{person}{Trevor Mudge}, {and}
  \bibinfo{person}{Richard~B Brown}.} \bibinfo{year}{2001}\natexlab{}.
\newblock \showarticletitle{MiBench: A free, commercially representative
  embedded benchmark suite}. In \bibinfo{booktitle}{\emph{Workload
  Characterization, 2001. WWC-4. 2001 IEEE International Workshop on}}. IEEE,
  \bibinfo{pages}{3--14}.
\newblock


\bibitem[\protect\citeauthoryear{Hester, Storer, Sitanayah, and Sorber}{Hester
  et~al\mbox{.}}{2016}]%
        {hester2016towards}
\bibfield{author}{\bibinfo{person}{Josiah Hester}, \bibinfo{person}{Kevin
  Storer}, \bibinfo{person}{Lanny Sitanayah}, {and} \bibinfo{person}{Jacob
  Sorber}.} \bibinfo{year}{2016}\natexlab{}.
\newblock \showarticletitle{Towards a Language and Runtime for
  Intermittently-Powered Devices}.
\newblock \bibinfo{journal}{\emph{sleep}}  \bibinfo{volume}{9}
  (\bibinfo{year}{2016}), \bibinfo{pages}{10}.
\newblock


\bibitem[\protect\citeauthoryear{Hicks}{Hicks}{2017}]%
        {Clank}
\bibfield{author}{\bibinfo{person}{Matthew Hicks}.}
  \bibinfo{year}{2017}\natexlab{}.
\newblock \showarticletitle{Clank: Architectural Support for Intermittent
  Computation}. In \bibinfo{booktitle}{\emph{In Proceedings of ISCA ’17}}.
  ACM.
\newblock


\bibitem[\protect\citeauthoryear{Hu, Ogleari, Zhao, Li, Basak, and Xie}{Hu
  et~al\mbox{.}}{2018}]%
        {hu2018persistence}
\bibfield{author}{\bibinfo{person}{Xing Hu}, \bibinfo{person}{Matheus Ogleari},
  \bibinfo{person}{Jishen Zhao}, \bibinfo{person}{Shuangchen Li},
  \bibinfo{person}{Abanti Basak}, {and} \bibinfo{person}{Yuan Xie}.}
  \bibinfo{year}{2018}\natexlab{}.
\newblock \showarticletitle{Persistence Parallelism Optimization: A Holistic
  Approach from Memory Bus to RDMA Network}. In
  \bibinfo{booktitle}{\emph{Proceedings of the 51st Annual IEEE/ACM
  International Symposium on Microarchitecture (MICRO)}}.
\newblock


\bibitem[\protect\citeauthoryear{Instruments}{Instruments}{2015}]%
        {msp430}
\bibfield{author}{\bibinfo{person}{Texas Instruments}.}
  \bibinfo{year}{2015}\natexlab{}.
\newblock \bibinfo{title}{MSP430FR family of Ultra Low-Power Microcontrollers}.
\newblock
\newblock
\newblock
\shownote{http://www.ti.com/lsds/ti/microcontrollers\_16-bit\_32-bit/msp/ultra-low\_power/msp430frxx\_fram/what\_is\_fram.page.}


\bibitem[\protect\citeauthoryear{Instruments}{Instruments}{2017}]%
        {msp430manual}
\bibfield{author}{\bibinfo{person}{Texas Instruments}.}
  \bibinfo{year}{2017}\natexlab{}.
\newblock \bibinfo{title}{MSP430FR59xx Mixed-Signal Microcontrollers}.
\newblock
\newblock
\newblock
\shownote{http://www.ti.com/lit/ds/symlink/msp430fr5969.pdf.}


\bibitem[\protect\citeauthoryear{Jayakumar, Lee, Lee, Raha, Kim, and
  Raghunathan}{Jayakumar et~al\mbox{.}}{2014a}]%
        {Jayakumar:2014}
\bibfield{author}{\bibinfo{person}{Hrishikesh Jayakumar},
  \bibinfo{person}{Kangwoo Lee}, \bibinfo{person}{Woo~Suk Lee},
  \bibinfo{person}{Arnab Raha}, \bibinfo{person}{Younghyun Kim}, {and}
  \bibinfo{person}{Vijay Raghunathan}.} \bibinfo{year}{2014}\natexlab{a}.
\newblock \showarticletitle{Powering the Internet of Things}. In
  \bibinfo{booktitle}{\emph{Proceedings of the 2014 International Symposium on
  Low Power Electronics and Design}} \emph{(\bibinfo{series}{ISLPED '14})}.
  \bibinfo{publisher}{ACM}, \bibinfo{address}{New York, NY, USA},
  \bibinfo{pages}{375--380}.
\newblock
\showISBNx{978-1-4503-2975-0}
\urldef\tempurl%
\url{https://doi.org/10.1145/2627369.2631644}
\showDOI{\tempurl}


\bibitem[\protect\citeauthoryear{Jayakumar, Raha, and Raghunathan}{Jayakumar
  et~al\mbox{.}}{2014b}]%
        {JayakumarRR14}
\bibfield{author}{\bibinfo{person}{Hrishikesh Jayakumar},
  \bibinfo{person}{Arnab Raha}, {and} \bibinfo{person}{Vijay Raghunathan}.}
  \bibinfo{year}{2014}\natexlab{b}.
\newblock \showarticletitle{QUICKRECALL: A Low Overhead HW/SW Approach for
  Enabling Computations across Power Cycles in Transiently Powered Computers.}.
  In \bibinfo{booktitle}{\emph{VLSI Design}}. \bibinfo{publisher}{IEEE Computer
  Society}, \bibinfo{pages}{330--335}.
\newblock
\showISBNx{978-1-4799-2513-1}
\urldef\tempurl%
\url{http://dblp.uni-trier.de/db/conf/vlsid/vlsid2014.html#JayakumarRR14}
\showURL{%
\tempurl}


\bibitem[\protect\citeauthoryear{Jayakumar, Raha, and Raghunathan}{Jayakumar
  et~al\mbox{.}}{2014c}]%
        {jayakumar2014quickrecall}
\bibfield{author}{\bibinfo{person}{Harishankar Jayakumar},
  \bibinfo{person}{Arnab Raha}, {and} \bibinfo{person}{Vijay Raghunathan}.}
  \bibinfo{year}{2014}\natexlab{c}.
\newblock \showarticletitle{QuickRecall: A low overhead HW/SW approach for
  enabling computations across power cycles in transiently powered computers}.
  In \bibinfo{booktitle}{\emph{VLSI Design and 2014 13th International
  Conference on Embedded Systems, 2014 27th International Conference on}}.
  IEEE, \bibinfo{pages}{330--335}.
\newblock


\bibitem[\protect\citeauthoryear{Jung, Lim, Lee, and Han}{Jung
  et~al\mbox{.}}{2005}]%
        {jung2005adaptive}
\bibfield{author}{\bibinfo{person}{Changhee Jung}, \bibinfo{person}{Daeseob
  Lim}, \bibinfo{person}{Jaejin Lee}, {and} \bibinfo{person}{SangYong Han}.}
  \bibinfo{year}{2005}\natexlab{}.
\newblock \showarticletitle{Adaptive execution techniques for SMT
  multiprocessor architectures}. In \bibinfo{booktitle}{\emph{Proceedings of
  the tenth ACM SIGPLAN symposium on Principles and practice of parallel
  programming}}. ACM, \bibinfo{pages}{236--246}.
\newblock


\bibitem[\protect\citeauthoryear{Kellogg, Talla, Gollakota, and Smith}{Kellogg
  et~al\mbox{.}}{2016}]%
        {kellogg2016passive}
\bibfield{author}{\bibinfo{person}{Bryce Kellogg}, \bibinfo{person}{Vamsi
  Talla}, \bibinfo{person}{Shyamnath Gollakota}, {and}
  \bibinfo{person}{Joshua~R Smith}.} \bibinfo{year}{2016}\natexlab{}.
\newblock \showarticletitle{Passive Wi-Fi: Bringing Low Power to Wi-Fi
  Transmissions.}. In \bibinfo{booktitle}{\emph{NSDI}},
  Vol.~\bibinfo{volume}{16}. \bibinfo{pages}{151--164}.
\newblock


\bibitem[\protect\citeauthoryear{Kim, Zeng, Liu, Adbel-Majeed, Lee, and
  Jung}{Kim et~al\mbox{.}}{2020}]%
        {kim2020penny}
\bibfield{author}{\bibinfo{person}{Hongjune Kim}, \bibinfo{person}{Jianping
  Zeng}, \bibinfo{person}{Qingrui Liu}, \bibinfo{person}{Mohammad
  Adbel-Majeed}, \bibinfo{person}{Jaejin Lee}, {and} \bibinfo{person}{Changhee
  Jung}.} \bibinfo{year}{2020}\natexlab{}.
\newblock \showarticletitle{Compiler-Directed Soft Error Resilience for
  Lightweight GPU Register File Protection}. In
  \bibinfo{booktitle}{\emph{Proceedings of the 41st ACM SIGPLAN Conference on
  Programming Language Design and Implementation}} \emph{(\bibinfo{series}{PLDI
  '20})}.
\newblock


\bibitem[\protect\citeauthoryear{Lattner and Adve}{Lattner and Adve}{2004}]%
        {lattner04llvm}
\bibfield{author}{\bibinfo{person}{Chris Lattner} {and} \bibinfo{person}{Vikram
  Adve}.} \bibinfo{year}{2004}\natexlab{}.
\newblock \showarticletitle{LLVM: A Compilation Framework for Lifelong Program
  Analysis \& Transformation}. In \bibinfo{booktitle}{\emph{Proceedings of the
  International Symposium on Code Generation and Optimization}}
  \emph{(\bibinfo{series}{CGO '04})}. \bibinfo{publisher}{IEEE Computer
  Society}, \bibinfo{address}{Washington, DC, USA}, \bibinfo{pages}{75--}.
\newblock


\bibitem[\protect\citeauthoryear{Lee, Potkonjak, and Mangione-Smith}{Lee
  et~al\mbox{.}}{1997}]%
        {lee1997mediabench}
\bibfield{author}{\bibinfo{person}{Chunho Lee}, \bibinfo{person}{Miodrag
  Potkonjak}, {and} \bibinfo{person}{William~H. Mangione-Smith}.}
  \bibinfo{year}{1997}\natexlab{}.
\newblock \showarticletitle{MediaBench: A Tool for Evaluating and Synthesizing
  Multimedia and Communicatons Systems}. In
  \bibinfo{booktitle}{\emph{Proceedings of the 30th Annual ACM/IEEE
  International Symposium on Microarchitecture}} \emph{(\bibinfo{series}{MICRO
  30})}. \bibinfo{publisher}{IEEE Computer Society},
  \bibinfo{address}{Washington, DC, USA}, \bibinfo{pages}{330--335}.
\newblock
\showISBNx{0-8186-7977-8}
\urldef\tempurl%
\url{http://dl.acm.org/citation.cfm?id=266800.266832}
\showURL{%
\tempurl}


\bibitem[\protect\citeauthoryear{Lee and Chang}{Lee and Chang}{2015}]%
        {lee2015powering}
\bibfield{author}{\bibinfo{person}{Hyung~Gyu Lee} {and}
  \bibinfo{person}{Naehyuck Chang}.} \bibinfo{year}{2015}\natexlab{}.
\newblock \showarticletitle{Powering the IoT: Storage-less and converter-less
  energy harvesting}. In \bibinfo{booktitle}{\emph{Design Automation Conference
  (ASP-DAC), 2015 20th Asia and South Pacific}}. IEEE,
  \bibinfo{pages}{124--129}.
\newblock


\bibitem[\protect\citeauthoryear{Lee, Park, Kim, Jung, Lim, and Han}{Lee
  et~al\mbox{.}}{2010}]%
        {adp10}
\bibfield{author}{\bibinfo{person}{Jaejin Lee}, \bibinfo{person}{Jung-Ho Park},
  \bibinfo{person}{Honggyu Kim}, \bibinfo{person}{Changhee Jung},
  \bibinfo{person}{Daeseob Lim}, {and} \bibinfo{person}{SangYong Han}.}
  \bibinfo{year}{2010}\natexlab{}.
\newblock \showarticletitle{Adaptive execution techniques of parallel programs
  for multiprocessors}.
\newblock \bibinfo{journal}{\emph{J. Parallel Distrib. Comput.}}
  \bibinfo{volume}{70}, \bibinfo{number}{5} (\bibinfo{date}{May}
  \bibinfo{year}{2010}), \bibinfo{pages}{467--480}.
\newblock
\showISSN{0743-7315}


\bibitem[\protect\citeauthoryear{Lee, Jayakumar, and Raghunathan}{Lee
  et~al\mbox{.}}{2014}]%
        {wslee14}
\bibfield{author}{\bibinfo{person}{Woo~Suk Lee}, \bibinfo{person}{Hrishikesh
  Jayakumar}, {and} \bibinfo{person}{Vijay Raghunathan}.}
  \bibinfo{year}{2014}\natexlab{}.
\newblock \showarticletitle{When they are not listening: Harvesting power from
  idle sensors in embedded systems}. In \bibinfo{booktitle}{\emph{Proceeding of
  the 5th International Green Computing Conference (IGCC),}}.
\newblock


\bibitem[\protect\citeauthoryear{Liu, Izraelevitz, Lee, Scott, Noh, and
  Jung}{Liu et~al\mbox{.}}{2018}]%
        {ido2018}
\bibfield{author}{\bibinfo{person}{Qingrui Liu}, \bibinfo{person}{Joseph
  Izraelevitz}, \bibinfo{person}{Se~Kwon Lee}, \bibinfo{person}{Michael~L.
  Scott}, \bibinfo{person}{Sam~H. Noh}, {and} \bibinfo{person}{Changhee Jung}.}
  \bibinfo{year}{2018}\natexlab{}.
\newblock \showarticletitle{iDO: Compiler-Directed Failure Atomicity for
  Nonvolatile Memory}. In \bibinfo{booktitle}{\emph{51st Annual {IEEE/ACM}
  International Symposium on Microarchitecture, {MICRO} 2018, Fukuoka, Japan,
  October 20-24, 2018}}. \bibinfo{pages}{258--270}.
\newblock
\urldef\tempurl%
\url{https://doi.org/10.1109/MICRO.2018.00029}
\showDOI{\tempurl}


\bibitem[\protect\citeauthoryear{Liu and Jung}{Liu and Jung}{2016}]%
        {TCCP}
\bibfield{author}{\bibinfo{person}{Qingrui Liu} {and} \bibinfo{person}{Changhee
  Jung}.} \bibinfo{year}{2016}\natexlab{}.
\newblock \showarticletitle{Lightweight hardware support for transparent
  consistency-aware checkpointing in intermittent energy-harvesting systems}.
  In \bibinfo{booktitle}{\emph{Non-Volatile Memory Systems and Applications
  Symposium (NVMSA), 2016 5th}}. IEEE, \bibinfo{pages}{1--6}.
\newblock


\bibitem[\protect\citeauthoryear{Liu, Jung, Lee, and Tiwari}{Liu
  et~al\mbox{.}}{2015a}]%
        {clover}
\bibfield{author}{\bibinfo{person}{Qingrui Liu}, \bibinfo{person}{Changhee
  Jung}, \bibinfo{person}{Dongyoon Lee}, {and} \bibinfo{person}{Devesh
  Tiwari}.} \bibinfo{year}{2015}\natexlab{a}.
\newblock \showarticletitle{Clover: Compiler Directed Lightweight Soft Error
  Resilience}. In \bibinfo{booktitle}{\emph{Proceedings of the 16th ACM
  SIGPLAN/SIGBED Conference on Languages, Compilers and Tools for Embedded
  Systems 2015 CD-ROM}} \emph{(\bibinfo{series}{LCTES'15})}.
  \bibinfo{publisher}{ACM}, \bibinfo{address}{New York, NY, USA}, Article
  \bibinfo{articleno}{2}, \bibinfo{numpages}{10}~pages.
\newblock
\showISBNx{978-1-4503-3257-6}
\urldef\tempurl%
\url{https://doi.org/10.1145/2670529.2754959}
\showDOI{\tempurl}


\bibitem[\protect\citeauthoryear{Liu, Jung, Lee, and Tiwari}{Liu
  et~al\mbox{.}}{2016a}]%
        {bolt}
\bibfield{author}{\bibinfo{person}{Qingrui Liu}, \bibinfo{person}{Changhee
  Jung}, \bibinfo{person}{Dongyoon Lee}, {and} \bibinfo{person}{Devesh
  Tiwari}.} \bibinfo{year}{2016}\natexlab{a}.
\newblock \showarticletitle{Compiler-directed lightweight checkpointing for
  fine-grained guaranteed soft error recovery}. In
  \bibinfo{booktitle}{\emph{High Performance Computing, Networking, Storage and
  Analysis, SC16: International Conference for}}. IEEE,
  \bibinfo{pages}{228--239}.
\newblock


\bibitem[\protect\citeauthoryear{Liu, Jung, Lee, and Tiwari}{Liu
  et~al\mbox{.}}{2016b}]%
        {16Tecs}
\bibfield{author}{\bibinfo{person}{Qingrui Liu}, \bibinfo{person}{Changhee
  Jung}, \bibinfo{person}{Dongyoon Lee}, {and} \bibinfo{person}{Devesh
  Tiwari}.} \bibinfo{year}{2016}\natexlab{b}.
\newblock \showarticletitle{Compiler-Directed Soft Error Detection and Recovery
  to Avoid DUE and SDC via Tail-DMR}.
\newblock \bibinfo{journal}{\emph{ACM Transactions on Embedded Computing
  Systems (TECS)}} \bibinfo{volume}{16}, \bibinfo{number}{2},
  \bibinfo{pages}{32}.
\newblock


\bibitem[\protect\citeauthoryear{Liu, Jung, Lee, and Tiwarit}{Liu
  et~al\mbox{.}}{2016c}]%
        {16Micro}
\bibfield{author}{\bibinfo{person}{Qingrui Liu}, \bibinfo{person}{Changhee
  Jung}, \bibinfo{person}{Dongyoon Lee}, {and} \bibinfo{person}{Devesh
  Tiwarit}.} \bibinfo{year}{2016}\natexlab{c}.
\newblock \showarticletitle{Low-cost soft error resilience with unified data
  verification and fine-grained recovery for acoustic sensor based detection}.
  In \bibinfo{booktitle}{\emph{Microarchitecture (MICRO), 2016 49th Annual
  IEEE/ACM International Symposium on}}. IEEE, \bibinfo{pages}{1--12}.
\newblock


\bibitem[\protect\citeauthoryear{Liu, Wu, Kittinger, Levy, and Jung}{Liu
  et~al\mbox{.}}{2017}]%
        {liu2017benchprime}
\bibfield{author}{\bibinfo{person}{Qingrui Liu}, \bibinfo{person}{Xiaolong Wu},
  \bibinfo{person}{Larry Kittinger}, \bibinfo{person}{Markus Levy}, {and}
  \bibinfo{person}{Changhee Jung}.} \bibinfo{year}{2017}\natexlab{}.
\newblock \showarticletitle{Benchprime: Effective building of a hybrid
  benchmark suite}.
\newblock \bibinfo{journal}{\emph{ACM Transactions on Embedded Computing
  Systems (TECS)}} \bibinfo{volume}{16}, \bibinfo{number}{5s}
  (\bibinfo{year}{2017}), \bibinfo{pages}{179}.
\newblock


\bibitem[\protect\citeauthoryear{Liu, Li, Li, Wang, Li, Ma, Li, Chang, John,
  Xie, et~al\mbox{.}}{Liu et~al\mbox{.}}{2015b}]%
        {Ambient}
\bibfield{author}{\bibinfo{person}{Yongpan Liu}, \bibinfo{person}{Zewei Li},
  \bibinfo{person}{Hehe Li}, \bibinfo{person}{Yiqun Wang},
  \bibinfo{person}{Xueqing Li}, \bibinfo{person}{Kaisheng Ma},
  \bibinfo{person}{Shuangchen Li}, \bibinfo{person}{Meng-Fan Chang},
  \bibinfo{person}{Sampson John}, \bibinfo{person}{Yuan Xie}, {et~al\mbox{.}}}
  \bibinfo{year}{2015}\natexlab{b}.
\newblock \showarticletitle{Ambient energy harvesting nonvolatile processors:
  from circuit to system}. In \bibinfo{booktitle}{\emph{Proceedings of the 52nd
  Annual Design Automation Conference}}. ACM, \bibinfo{pages}{150}.
\newblock


\bibitem[\protect\citeauthoryear{Lucia, Balaji, Colin, Maeng, and Ruppel}{Lucia
  et~al\mbox{.}}{2017}]%
        {lucia2017intermittent}
\bibfield{author}{\bibinfo{person}{Brandon Lucia}, \bibinfo{person}{Vignesh
  Balaji}, \bibinfo{person}{Alexei Colin}, \bibinfo{person}{Kiwan Maeng}, {and}
  \bibinfo{person}{Emily Ruppel}.} \bibinfo{year}{2017}\natexlab{}.
\newblock \showarticletitle{Intermittent Computing: Challenges and
  Opportunities}. In \bibinfo{booktitle}{\emph{LIPIcs-Leibniz International
  Proceedings in Informatics}}, Vol.~\bibinfo{volume}{71}. Schloss
  Dagstuhl-Leibniz-Zentrum fuer Informatik.
\newblock


\bibitem[\protect\citeauthoryear{Lucia and Ransford}{Lucia and
  Ransford}{2015}]%
        {dino}
\bibfield{author}{\bibinfo{person}{Brandon Lucia} {and}
  \bibinfo{person}{Benjamin Ransford}.} \bibinfo{year}{2015}\natexlab{}.
\newblock \showarticletitle{A simpler, safer programming and execution model
  for intermittent systems}. In \bibinfo{booktitle}{\emph{Proceedings of the
  36th ACM SIGPLAN Conference on Programming Language Design and
  Implementation}}. ACM, \bibinfo{pages}{575--585}.
\newblock


\bibitem[\protect\citeauthoryear{Lui, Li, Li, Wang, Li, Ma, Li, Chang, Sampson,
  Xie, Shu, and Yang}{Lui et~al\mbox{.}}{2015}]%
        {Lui_2015}
\bibfield{author}{\bibinfo{person}{Yongpan Lui}, \bibinfo{person}{Zewel Li},
  \bibinfo{person}{Hehe Li}, \bibinfo{person}{Yiqun Wang},
  \bibinfo{person}{Xueqing Li}, \bibinfo{person}{Kaisheng Ma},
  \bibinfo{person}{Shuangchen Li}, \bibinfo{person}{Meng-Fan Chang},
  \bibinfo{person}{Jack Sampson}, \bibinfo{person}{Yuan Xie},
  \bibinfo{person}{Jiwu Shu}, {and} \bibinfo{person}{Huazhong Yang}.}
  \bibinfo{year}{2015}\natexlab{}.
\newblock \showarticletitle{Ambient Energy Harvesting Nonvolatile Processors:
  From Circuit to System}. In \bibinfo{booktitle}{\emph{Proceedings of the 52nd
  Annual Design Automation Conference}} \emph{(\bibinfo{series}{DAC '15})}.
  \bibinfo{publisher}{ACM}, \bibinfo{address}{New York, NY, USA}, 6.
\newblock
\showISBNx{978-1-4503-3520-1}


\bibitem[\protect\citeauthoryear{Ma, Li, Li, Liu, Xie, Sampson, Kandemir, and
  Narayanan}{Ma et~al\mbox{.}}{2017a}]%
        {ma2017incidental}
\bibfield{author}{\bibinfo{person}{Kaisheng Ma}, \bibinfo{person}{Xueqing Li},
  \bibinfo{person}{Jinyang Li}, \bibinfo{person}{Yongpwan Liu},
  \bibinfo{person}{Yuan Xie}, \bibinfo{person}{Jack Sampson},
  \bibinfo{person}{Mahmut~Taylan Kandemir}, {and}
  \bibinfo{person}{Vijaykrishnan Narayanan}.} \bibinfo{year}{2017}\natexlab{a}.
\newblock \showarticletitle{Incidental computing on IoT nonvolatile
  processors}. In \bibinfo{booktitle}{\emph{Proceedings of the 50th Annual
  IEEE/ACM International Symposium on Microarchitecture}}. ACM,
  \bibinfo{pages}{204--218}.
\newblock


\bibitem[\protect\citeauthoryear{Ma, Li, Srinivasa, Liu, Sampson, Xie, and
  Narayanan}{Ma et~al\mbox{.}}{2017b}]%
        {ma2017spendthrift}
\bibfield{author}{\bibinfo{person}{Kaisheng Ma}, \bibinfo{person}{Xueqing Li},
  \bibinfo{person}{Srivatsa~Rangachar Srinivasa}, \bibinfo{person}{Yongpan
  Liu}, \bibinfo{person}{John Sampson}, \bibinfo{person}{Yuan Xie}, {and}
  \bibinfo{person}{Vijaykrishnan Narayanan}.} \bibinfo{year}{2017}\natexlab{b}.
\newblock \showarticletitle{Spendthrift: Machine learning based resource and
  frequency scaling for ambient energy harvesting nonvolatile processors}. In
  \bibinfo{booktitle}{\emph{Design Automation Conference (ASP-DAC), 2017 22nd
  Asia and South Pacific}}. IEEE, \bibinfo{pages}{678--683}.
\newblock


\bibitem[\protect\citeauthoryear{Ma, Zheng, Li, Swaminathan, Li, Liu, Sampson,
  Xie, and Narayanan}{Ma et~al\mbox{.}}{2015}]%
        {ma2015architecture}
\bibfield{author}{\bibinfo{person}{Kaisheng Ma}, \bibinfo{person}{Yang Zheng},
  \bibinfo{person}{Shuangchen Li}, \bibinfo{person}{Karthik Swaminathan},
  \bibinfo{person}{Xueqing Li}, \bibinfo{person}{Yongpan Liu},
  \bibinfo{person}{Jack Sampson}, \bibinfo{person}{Yuan Xie}, {and}
  \bibinfo{person}{Vijaykrishnan Narayanan}.} \bibinfo{year}{2015}\natexlab{}.
\newblock \showarticletitle{Architecture exploration for ambient energy
  harvesting nonvolatile processors}. In \bibinfo{booktitle}{\emph{High
  Performance Computer Architecture (HPCA), 2015 IEEE 21st International
  Symposium on}}. IEEE, \bibinfo{pages}{526--537}.
\newblock


\bibitem[\protect\citeauthoryear{MAENG, COLIN, and LUCIA}{MAENG
  et~al\mbox{.}}{2017}]%
        {maeng2017alpaca}
\bibfield{author}{\bibinfo{person}{KIWAN MAENG}, \bibinfo{person}{ALEXEI
  COLIN}, {and} \bibinfo{person}{BRANDON LUCIA}.}
  \bibinfo{year}{2017}\natexlab{}.
\newblock \showarticletitle{Alpaca: Intermittent Execution without
  Checkpoints}. In \bibinfo{booktitle}{\emph{Proceedings of the 2016 ACM
  SIGPLAN International Conference on Object-Oriented Programming, Systems,
  Languages, and Applications}}. ACM.
\newblock


\bibitem[\protect\citeauthoryear{Maeng and Lucia}{Maeng and Lucia}{2018}]%
        {Chinchilla}
\bibfield{author}{\bibinfo{person}{Kiwan Maeng} {and} \bibinfo{person}{Brandon
  Lucia}.} \bibinfo{year}{2018}\natexlab{}.
\newblock \showarticletitle{Adaptive Dynamic Checkpointing for Safe Efficient
  Intermittent Computing}. In \bibinfo{booktitle}{\emph{13th {USENIX} Symposium
  on Operating Systems Design and Implementation ({OSDI} 18)}}.
  \bibinfo{publisher}{{USENIX} Association}, \bibinfo{address}{Carlsbad, CA},
  \bibinfo{pages}{129--144}.
\newblock
\showISBNx{978-1-931971-47-8}
\urldef\tempurl%
\url{https://www.usenix.org/conference/osdi18/presentation/maeng}
\showURL{%
\tempurl}


\bibitem[\protect\citeauthoryear{Maeng and Lucia}{Maeng and Lucia}{2019}]%
        {maeng2019supporting}
\bibfield{author}{\bibinfo{person}{Kiwan Maeng} {and} \bibinfo{person}{Brandon
  Lucia}.} \bibinfo{year}{2019}\natexlab{}.
\newblock \showarticletitle{Supporting peripherals in intermittent systems with
  just-in-time checkpoints}. In \bibinfo{booktitle}{\emph{Proceedings of the
  40th ACM SIGPLAN Conference on Programming Language Design and
  Implementation}}. ACM, \bibinfo{pages}{1101--1116}.
\newblock


\bibitem[\protect\citeauthoryear{Mehta and Torrellas}{Mehta and
  Torrellas}{2016}]%
        {mehta2016wearcore}
\bibfield{author}{\bibinfo{person}{Sanyam Mehta} {and} \bibinfo{person}{Josep
  Torrellas}.} \bibinfo{year}{2016}\natexlab{}.
\newblock \showarticletitle{WearCore: A core for wearable workloads?}. In
  \bibinfo{booktitle}{\emph{Parallel Architecture and Compilation Techniques
  (PACT), 2016 International Conference on}}. IEEE, \bibinfo{pages}{153--164}.
\newblock


\bibitem[\protect\citeauthoryear{Naderiparizi, Parks, Kapetanovic, Ransford,
  and Smith}{Naderiparizi et~al\mbox{.}}{2015}]%
        {WISPcam}
\bibfield{author}{\bibinfo{person}{Saman Naderiparizi},
  \bibinfo{person}{Aaron~N Parks}, \bibinfo{person}{Zerina Kapetanovic},
  \bibinfo{person}{Benjamin Ransford}, {and} \bibinfo{person}{Joshua~R Smith}.}
  \bibinfo{year}{2015}\natexlab{}.
\newblock \showarticletitle{Wispcam: A battery-free rfid camera}. In
  \bibinfo{booktitle}{\emph{RFID (RFID), 2015 IEEE International Conference
  on}}. IEEE, \bibinfo{pages}{166--173}.
\newblock


\bibitem[\protect\citeauthoryear{Nair, Chou, Rajendran, and Qureshi}{Nair
  et~al\mbox{.}}{2015}]%
        {nair2015reducing}
\bibfield{author}{\bibinfo{person}{Prashant~J Nair}, \bibinfo{person}{Chiachen
  Chou}, \bibinfo{person}{Bipin Rajendran}, {and} \bibinfo{person}{Moinuddin~K
  Qureshi}.} \bibinfo{year}{2015}\natexlab{}.
\newblock \showarticletitle{Reducing read latency of phase change memory via
  early read and turbo read}. In \bibinfo{booktitle}{\emph{2015 IEEE 21st
  International Symposium on High Performance Computer Architecture (HPCA)}}.
  IEEE, \bibinfo{pages}{309--319}.
\newblock


\bibitem[\protect\citeauthoryear{Nirjon}{Nirjon}{2018}]%
        {nirjon2018lifelong}
\bibfield{author}{\bibinfo{person}{Shahriar Nirjon}.}
  \bibinfo{year}{2018}\natexlab{}.
\newblock \showarticletitle{Lifelong Learning on Harvested Energy}. In
  \bibinfo{booktitle}{\emph{Proceedings of the 16th Annual International
  Conference on Mobile Systems, Applications, and Services}}. ACM,
  \bibinfo{pages}{500--501}.
\newblock


\bibitem[\protect\citeauthoryear{Nwafor, Campbell, Hill, and Bloom}{Nwafor
  et~al\mbox{.}}{2017}]%
        {nwafor2017towards}
\bibfield{author}{\bibinfo{person}{Ebelechukwu Nwafor}, \bibinfo{person}{Andre
  Campbell}, \bibinfo{person}{David Hill}, {and} \bibinfo{person}{Gedare
  Bloom}.} \bibinfo{year}{2017}\natexlab{}.
\newblock \showarticletitle{Towards a provenance collection framework for
  Internet of Things devices}. In \bibinfo{booktitle}{\emph{2017 IEEE
  SmartWorld, Ubiquitous Intelligence \& Computing, Advanced \& Trusted
  Computed, Scalable Computing \& Communications, Cloud \& Big Data Computing,
  Internet of People and Smart City Innovation
  (SmartWorld/SCALCOM/UIC/ATC/CBDCom/IOP/SCI)}}. IEEE, \bibinfo{pages}{1--6}.
\newblock


\bibitem[\protect\citeauthoryear{Rizzon, Rossi, Passerone, and Brunelli}{Rizzon
  et~al\mbox{.}}{2013}]%
        {Rizzon:2013}
\bibfield{author}{\bibinfo{person}{Luca Rizzon}, \bibinfo{person}{Maurizio
  Rossi}, \bibinfo{person}{Roberto Passerone}, {and} \bibinfo{person}{Davide
  Brunelli}.} \bibinfo{year}{2013}\natexlab{}.
\newblock \showarticletitle{Wireless Sensor Networks for Environmental
  Monitoring Powered by Microprocessors Heat Dissipation}. In
  \bibinfo{booktitle}{\emph{Proceedings of the 1st International Workshop on
  Energy Neutral Sensing Systems}} \emph{(\bibinfo{series}{ENSSys '13})}.
  \bibinfo{publisher}{ACM}, \bibinfo{address}{New York, NY, USA}, Article
  \bibinfo{articleno}{8}, \bibinfo{numpages}{6}~pages.
\newblock
\showISBNx{978-1-4503-2432-8}
\urldef\tempurl%
\url{https://doi.org/10.1145/2534208.2534216}
\showDOI{\tempurl}


\bibitem[\protect\citeauthoryear{Rodriguez~Arreola, Balsamo, Das, Weddell,
  Brunelli, Al-Hashimi, and Merrett}{Rodriguez~Arreola et~al\mbox{.}}{2015}]%
        {hibernus2015}
\bibfield{author}{\bibinfo{person}{Alberto Rodriguez~Arreola},
  \bibinfo{person}{Domenico Balsamo}, \bibinfo{person}{Anup~K. Das},
  \bibinfo{person}{Alex~S. Weddell}, \bibinfo{person}{Davide Brunelli},
  \bibinfo{person}{Bashir~M. Al-Hashimi}, {and} \bibinfo{person}{Geoff~V.
  Merrett}.} \bibinfo{year}{2015}\natexlab{}.
\newblock \showarticletitle{Approaches to Transient Computing for Energy
  Harvesting Systems: A Quantitative Evaluation}. In
  \bibinfo{booktitle}{\emph{Proceedings of the 3rd International Workshop on
  Energy Harvesting \&\#38; Energy Neutral Sensing Systems}}
  \emph{(\bibinfo{series}{ENSsys '15})}. \bibinfo{publisher}{ACM},
  \bibinfo{address}{New York, NY, USA}, \bibinfo{pages}{3--8}.
\newblock
\showISBNx{978-1-4503-3837-0}
\urldef\tempurl%
\url{https://doi.org/10.1145/2820645.2820652}
\showDOI{\tempurl}


\bibitem[\protect\citeauthoryear{Ruppel and Lucia}{Ruppel and Lucia}{2019}]%
        {ruppel2019transactional}
\bibfield{author}{\bibinfo{person}{Emily Ruppel} {and} \bibinfo{person}{Brandon
  Lucia}.} \bibinfo{year}{2019}\natexlab{}.
\newblock \showarticletitle{Transactional concurrency control for intermittent,
  energy-harvesting computing systems}. In
  \bibinfo{booktitle}{\emph{Proceedings of the 40th ACM SIGPLAN Conference on
  Programming Language Design and Implementation}}. ACM,
  \bibinfo{pages}{1085--1100}.
\newblock


\bibitem[\protect\citeauthoryear{Shigeta, Sasaki, Quan, Kawahara, Vyas,
  Tentzeris, and Asami}{Shigeta et~al\mbox{.}}{2013}]%
        {CapLeak}
\bibfield{author}{\bibinfo{person}{Ryo Shigeta}, \bibinfo{person}{Tatsuya
  Sasaki}, \bibinfo{person}{Duong~Minh Quan}, \bibinfo{person}{Yoshihiro
  Kawahara}, \bibinfo{person}{Rushi~J Vyas}, \bibinfo{person}{Manos~M
  Tentzeris}, {and} \bibinfo{person}{Tohru Asami}.}
  \bibinfo{year}{2013}\natexlab{}.
\newblock \showarticletitle{Ambient RF energy harvesting sensor device with
  capacitor-leakage-aware duty cycle control}.
\newblock \bibinfo{journal}{\emph{IEEE Sensors Journal}} \bibinfo{volume}{13},
  \bibinfo{number}{8}, \bibinfo{pages}{2973--2983}.
\newblock


\bibitem[\protect\citeauthoryear{Shivakumar and Jouppi}{Shivakumar and
  Jouppi}{2001}]%
        {shivakumar2001cacti}
\bibfield{author}{\bibinfo{person}{Premkishore Shivakumar} {and}
  \bibinfo{person}{Norman~P Jouppi}.} \bibinfo{year}{2001}\natexlab{}.
\newblock \showarticletitle{Cacti 3.0: An integrated cache timing, power, and
  area model}.
\newblock  (\bibinfo{year}{2001}).
\newblock


\bibitem[\protect\citeauthoryear{Sisinni, Saifullah, Han, Jennehag, and
  Gidlund}{Sisinni et~al\mbox{.}}{2018}]%
        {sisinni2018industrial}
\bibfield{author}{\bibinfo{person}{Emiliano Sisinni},
  \bibinfo{person}{Abusayeed Saifullah}, \bibinfo{person}{Song Han},
  \bibinfo{person}{Ulf Jennehag}, {and} \bibinfo{person}{Mikael Gidlund}.}
  \bibinfo{year}{2018}\natexlab{}.
\newblock \showarticletitle{Industrial Internet of Things: Challenges,
  Opportunities, and Directions}.
\newblock \bibinfo{journal}{\emph{IEEE Transactions on Industrial Informatics}}
  (\bibinfo{year}{2018}).
\newblock


\bibitem[\protect\citeauthoryear{Smith}{Smith}{2013}]%
        {WISP}
\bibfield{author}{\bibinfo{person}{Joshua~R. Smith}.}
  \bibinfo{year}{2013}\natexlab{}.
\newblock \bibinfo{booktitle}{\emph{Wirelessly Powered Sensor Networks and
  Computational RFID}}.
\newblock \bibinfo{publisher}{Springer}, \bibinfo{address}{New York, NY, USA}.
\newblock


\bibitem[\protect\citeauthoryear{Su, Liu, Wang, and Yang}{Su
  et~al\mbox{.}}{2017}]%
        {su2017ferroelectric}
\bibfield{author}{\bibinfo{person}{Fang Su}, \bibinfo{person}{Yongpan Liu},
  \bibinfo{person}{Yiqun Wang}, {and} \bibinfo{person}{Huazhong Yang}.}
  \bibinfo{year}{2017}\natexlab{}.
\newblock \showarticletitle{A Ferroelectric Nonvolatile Processor with 46$\mu$
  s System-Level Wake-up Time and 14$\mu$ s Sleep Time for Energy Harvesting
  Applications}.
\newblock \bibinfo{journal}{\emph{IEEE Transactions on Circuits and Systems I:
  Regular Papers}} \bibinfo{volume}{64}, \bibinfo{number}{3}
  (\bibinfo{year}{2017}), \bibinfo{pages}{596--607}.
\newblock


\bibitem[\protect\citeauthoryear{Sui and Xue}{Sui and Xue}{2016}]%
        {sui2016svf}
\bibfield{author}{\bibinfo{person}{Yulei Sui} {and} \bibinfo{person}{Jingling
  Xue}.} \bibinfo{year}{2016}\natexlab{}.
\newblock \showarticletitle{SVF: interprocedural static value-flow analysis in
  LLVM}. In \bibinfo{booktitle}{\emph{Proceedings of the 25th International
  Conference on Compiler Construction}}. ACM, \bibinfo{pages}{265--266}.
\newblock


\bibitem[\protect\citeauthoryear{Teverovsky}{Teverovsky}{2014}]%
        {capcrack}
\bibfield{author}{\bibinfo{person}{Alexander Teverovsky}.}
  \bibinfo{year}{2014}\natexlab{}.
\newblock \showarticletitle{Insulation resistance and leakage currents in
  low-voltage ceramic capacitors with cracks}.
\newblock \bibinfo{journal}{\emph{IEEE Transactions on Components, Packaging
  and Manufacturing Technology}} \bibinfo{volume}{4}, \bibinfo{number}{7}
  (\bibinfo{year}{2014}), \bibinfo{pages}{1169--1176}.
\newblock


\bibitem[\protect\citeauthoryear{Vedula, Shriraman, Kumar, and Sumner}{Vedula
  et~al\mbox{.}}{2018}]%
        {vedula2018nachos}
\bibfield{author}{\bibinfo{person}{Naveen Vedula}, \bibinfo{person}{Arrvindh
  Shriraman}, \bibinfo{person}{Snehasish Kumar}, {and}
  \bibinfo{person}{William~N Sumner}.} \bibinfo{year}{2018}\natexlab{}.
\newblock \showarticletitle{NACHOS: Software-Driven Hardware-Assisted Memory
  Disambiguation for Accelerators}. In \bibinfo{booktitle}{\emph{2018 IEEE
  International Symposium on High Performance Computer Architecture (HPCA)}}.
  IEEE, \bibinfo{pages}{710--723}.
\newblock


\bibitem[\protect\citeauthoryear{Wang, Chang, Kim, Park, Liu, Lee, Luo, and
  Yang}{Wang et~al\mbox{.}}{2014}]%
        {hglee14}
\bibfield{author}{\bibinfo{person}{Cong Wang}, \bibinfo{person}{Naehyuck
  Chang}, \bibinfo{person}{Younghyun Kim}, \bibinfo{person}{Sangyoung Park},
  \bibinfo{person}{Yongpan Liu}, \bibinfo{person}{Hyung~Gyu Lee},
  \bibinfo{person}{Rong Luo}, {and} \bibinfo{person}{Huazhong Yang}.}
  \bibinfo{year}{2014}\natexlab{}.
\newblock \showarticletitle{Storage-less and converter-less maximum power point
  tracking of photovoltaic cells for a nonvolatile microprocessor}. In
  \bibinfo{booktitle}{\emph{Design Automation Conference (ASP-DAC), 2014 19th
  Asia and South Pacific}}. \bibinfo{pages}{379--384}.
\newblock
\urldef\tempurl%
\url{https://doi.org/10.1109/ASPDAC.2014.6742919}
\showDOI{\tempurl}


\bibitem[\protect\citeauthoryear{Wang and Wu}{Wang and Wu}{2013}]%
        {wang2013tso_atomicity}
\bibfield{author}{\bibinfo{person}{Cheng Wang} {and} \bibinfo{person}{Youfeng
  Wu}.} \bibinfo{year}{2013}\natexlab{}.
\newblock \showarticletitle{TSO\_ATOMICITY: efficient hardware primitive for
  TSO-preserving region optimizations}.
\newblock \bibinfo{journal}{\emph{ACM SIGPLAN Notices}} \bibinfo{volume}{48},
  \bibinfo{number}{4} (\bibinfo{year}{2013}), \bibinfo{pages}{509--520}.
\newblock


\bibitem[\protect\citeauthoryear{Wang, Liu, Li, Zhang, Zhao, Chiang, Yan, Sai,
  and Yang}{Wang et~al\mbox{.}}{2012}]%
        {3us}
\bibfield{author}{\bibinfo{person}{Yiqun Wang}, \bibinfo{person}{Yongpan Liu},
  \bibinfo{person}{Shuangchen Li}, \bibinfo{person}{Daming Zhang},
  \bibinfo{person}{Bo Zhao}, \bibinfo{person}{Mei-Fang Chiang},
  \bibinfo{person}{Yanxin Yan}, \bibinfo{person}{Baiko Sai}, {and}
  \bibinfo{person}{Huazhong Yang}.} \bibinfo{year}{2012}\natexlab{}.
\newblock \showarticletitle{A 3us wake-up time nonvolatile processor based on
  ferroelectric flip-flops}. In \bibinfo{booktitle}{\emph{ESSCIRC (ESSCIRC),
  2012 Proceedings of the}}. IEEE, \bibinfo{pages}{149--152}.
\newblock


\bibitem[\protect\citeauthoryear{Wang, Liu, Wang, Li, Sheng, Lee, Chang, and
  Yang}{Wang et~al\mbox{.}}{2016}]%
        {wang2016storage}
\bibfield{author}{\bibinfo{person}{Yiqun Wang}, \bibinfo{person}{Yongpan Liu},
  \bibinfo{person}{Cong Wang}, \bibinfo{person}{Zewei Li},
  \bibinfo{person}{Xiao Sheng}, \bibinfo{person}{Hyung~Gyu Lee},
  \bibinfo{person}{Naehyuck Chang}, {and} \bibinfo{person}{Huazhong Yang}.}
  \bibinfo{year}{2016}\natexlab{}.
\newblock \showarticletitle{Storage-less and converter-less photovoltaic energy
  harvesting with maximum power point tracking for internet of things}.
\newblock \bibinfo{journal}{\emph{IEEE Transactions on Computer-Aided Design of
  Integrated Circuits and Systems}} \bibinfo{volume}{35}, \bibinfo{number}{2}
  (\bibinfo{year}{2016}), \bibinfo{pages}{173--186}.
\newblock


\bibitem[\protect\citeauthoryear{Woude and Hicks}{Woude and Hicks}{2016}]%
        {Ratchet}
\bibfield{author}{\bibinfo{person}{Joel Van~Der Woude} {and}
  \bibinfo{person}{Matthew Hicks}.} \bibinfo{year}{2016}\natexlab{}.
\newblock \showarticletitle{Intermittent Computation without Hardware Support
  or Programmer Intervention}. In \bibinfo{booktitle}{\emph{12th {USENIX}
  Symposium on Operating Systems Design and Implementation ({OSDI} 16)}}.
  \bibinfo{publisher}{{USENIX} Association}, \bibinfo{address}{Savannah, GA},
  \bibinfo{pages}{17--32}.
\newblock
\showISBNx{978-1-931971-33-1}
\urldef\tempurl%
\url{https://www.usenix.org/conference/osdi16/technical-sessions/presentation/vanderwoude}
\showURL{%
\tempurl}


\bibitem[\protect\citeauthoryear{Xie, Zhao, Pan, Hu, Liu, and Xue}{Xie
  et~al\mbox{.}}{2015}]%
        {Xie_2015}
\bibfield{author}{\bibinfo{person}{Mimi Xie}, \bibinfo{person}{Mengying Zhao},
  \bibinfo{person}{Chao Pan}, \bibinfo{person}{Jingtong Hu},
  \bibinfo{person}{Yongpan Liu}, {and} \bibinfo{person}{Chun Xue}.}
  \bibinfo{year}{2015}\natexlab{}.
\newblock \showarticletitle{Fixing the Broken Time Machine: Consistency-Aware
  Checkpointing for Energy Harvesting Powered Non-Volatile Processor}. In
  \bibinfo{booktitle}{\emph{Proceedings of The 52nd IEEE/ACM Design Automation
  Conference (DAC 2015)}} \emph{(\bibinfo{series}{DAC '15})}.
  \bibinfo{publisher}{ACM}, \bibinfo{address}{New York, NY, USA}, 6.
\newblock


\bibitem[\protect\citeauthoryear{Xie}{Xie}{2016}]%
        {xie2016emerging}
\bibfield{author}{\bibinfo{person}{Yuan Xie}.} \bibinfo{year}{2016}\natexlab{}.
\newblock \bibinfo{booktitle}{\emph{EMERGING MEMORY TECHNOLOGIES.}}
\newblock \bibinfo{publisher}{Springer}.
\newblock


\bibitem[\protect\citeauthoryear{Xu, Niu, Muralimanohar, Balasubramonian,
  Zhang, Yu, and Xie}{Xu et~al\mbox{.}}{2015}]%
        {xu2015overcoming}
\bibfield{author}{\bibinfo{person}{Cong Xu}, \bibinfo{person}{Dimin Niu},
  \bibinfo{person}{Naveen Muralimanohar}, \bibinfo{person}{Rajeev
  Balasubramonian}, \bibinfo{person}{Tao Zhang}, \bibinfo{person}{Shimeng Yu},
  {and} \bibinfo{person}{Yuan Xie}.} \bibinfo{year}{2015}\natexlab{}.
\newblock \showarticletitle{Overcoming the challenges of crossbar resistive
  memory architectures}. In \bibinfo{booktitle}{\emph{2015 IEEE 21st
  International Symposium on High Performance Computer Architecture (HPCA)}}.
  IEEE, \bibinfo{pages}{476--488}.
\newblock


\bibitem[\protect\citeauthoryear{Y{\i}ld{\i}r{\i}m, Aantjes, Majid, and
  Pawe{\l}czak}{Y{\i}ld{\i}r{\i}m et~al\mbox{.}}{2016}]%
        {yildirim2016synchronization}
\bibfield{author}{\bibinfo{person}{Kas{\i}m~Sinan Y{\i}ld{\i}r{\i}m},
  \bibinfo{person}{Henko Aantjes}, \bibinfo{person}{Amjad~Yousef Majid}, {and}
  \bibinfo{person}{Przemys{\l}aw Pawe{\l}czak}.}
  \bibinfo{year}{2016}\natexlab{}.
\newblock \showarticletitle{On the synchronization of intermittently powered
  wireless embedded systems}.
\newblock \bibinfo{journal}{\emph{arXiv preprint arXiv:1606.01719}}
  (\bibinfo{year}{2016}).
\newblock


\end{thebibliography}



\ignore{
\appendix

\section{Research Methods}

\subsection{Part One}

Lorem ipsum dolor sit amet, consectetur adipiscing elit. Morbi
malesuada, quam in pulvinar varius, metus nunc fermentum urna, id
sollicitudin purus odio sit amet enim. Aliquam ullamcorper eu ipsum
vel mollis. Curabitur quis dictum nisl. Phasellus vel semper risus, et
lacinia dolor. Integer ultricies commodo sem nec semper.

\subsection{Part Two}

Etiam commodo feugiat nisl pulvinar pellentesque. Etiam auctor sodales
ligula, non varius nibh pulvinar semper. Suspendisse nec lectus non
ipsum convallis congue hendrerit vitae sapien. Donec at laoreet
eros. Vivamus non purus placerat, scelerisque diam eu, cursus
ante. Etiam aliquam tortor auctor efficitur mattis.

\section{Online Resources}

Nam id fermentum dui. Suspendisse sagittis tortor a nulla mollis, in
pulvinar ex pretium. Sed interdum orci quis metus euismod, et sagittis
enim maximus. Vestibulum gravida massa ut felis suscipit
congue. Quisque mattis elit a risus ultrices commodo venenatis eget
dui. Etiam sagittis eleifend elementum.

Nam interdum magna at lectus dignissim, ac dignissim lorem
rhoncus. Maecenas eu arcu ac neque placerat aliquam. Nunc pulvinar
massa et mattis lacinia.
}
\end{document}